\documentclass{article}
\usepackage{graphicx}
\usepackage[english]{babel}
\usepackage{epsfig}    
\usepackage[noadjust]{cite}
\title{\includegraphics[width=0.35\textwidth]{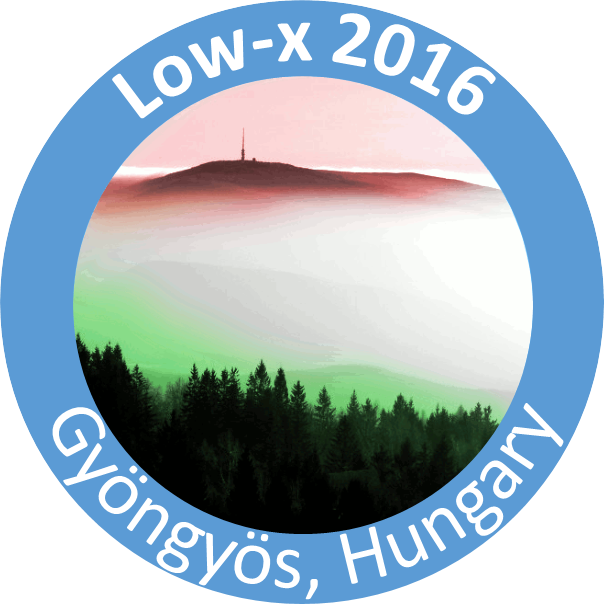}\\[1cm]
}

\title{
\includegraphics[width=0.35\textwidth]{lowx16-logo.png}\\[1cm]
Measurement of the underlying-event properties with the ATLAS detector
}
\author{{Yuri Kulchitsky$^{1, 2}$}\\[1ex]
on behalf of the ATLAS collaboration\\[3mm]
$^1$Institute of Physics, National Academy of Sciences, Minsk, Belarus\\
$^2$JINR, Dubna, Russia\\[3mm]
}
\begin{document}   
\fontfamily{lmss}\selectfont
\maketitle

%\linenumbers

\begin{abstract}
\noindent
 A correct modelling of the underlying event in  proton-proton
 collisions is important for the proper simulation of kinematic distributions of  high-energy collisions.
 The ATLAS collaboration extended previous studies at 7~TeV with a leading track or jet or Z boson by a new study of 
 Drell-Yan  events in 1.1~fb$^{-1}$ of data collected at a   center-of-mass  energy of 7~TeV.
 In this new study the distributions of several topological event-shape variables based on charged particles are measured, 
 both integrated and differential in the transverse momentum of the Drell-Yan lepton pair.
 These measurements are sensitive to the underlying-event as well as the onset of hard emissions. 
 The results have been compared with the predictions of several  state-of-the-art MC generators.
 The collaboration has also performed a first study of the number and  transverse momentum
 sum of charged particles as a function of  transverse momentum and azimuthal angle in a special 
 data set taken with low beam currents at a  center-of-mass energy of 13~TeV.
 The results are compared to predictions of several MC generators. 
\end{abstract}

%\vspace*{-5mm}
%_______________________________________________________________________
\section{Introduction}
\label{intro}
%_______________________________________________________________________
%
%
The Large Hadron Collider (LHC) was primarily built to explore the mechanism of electroweak symmetry breaking
and to search for new physics beyond the Standard Model (SM) in proton-proton collisions characterised by 
parton-parton scatterings with a high momentum transfer.  
These parton-parton scatterings are unavoidably accompanied by interactions between the proton remnants which are often
called the “underlying event” (UE) and have to be modelled well in order to be able to measure high-momentum-transfer
processes to high accuracy.
Since the UE is dominated by low-scale strong-force interactions, in which the strong coupling strength diverges and
perturbative methods of quantum chromodynamics (QCD) lose predictivity, it is extremely difficult to predict 
UE-sensitive observables from an ab-initio calculation in QCD.
As a result, one has to rely on models implemented in general-purpose Monte Carlo (MC) event generators. 
Generators such as Herwig~7 \cite{UE_shape_1}, Pythia~8 \cite{UE_shape_2}, Sherpa \cite{UE_shape_3}, and EPOS \cite{epos} 
contain multiple partonic interactions (MPI) as well as QCD radiation in the initial and final state to describe the UE. 
A correct modelling of the underlying event in  pp-collisions is  important 
for the proper simulation of kinematic distributions of   high-energy collisions.

Event-shape observables measured with the ATLAS detector \cite{ATLAS} at the LHC 
using charged particles in inclusive Z-boson events are presented in Ref.\ \cite{Aad:2016ria}.
The measurements are based on an integrated luminosity of 1.1~fb$^{-1}$
of pp-collisions at a centre-of-mass energy of $\sqrt{s} = 7$~TeV. 
Charged-particle distributions, excluding the lepton-antilepton pair from the Z-boson decay, 
are measured in different ranges of transverse momentum of the Z boson. 
These distributions include multiplicity, scalar sum of transverse momenta, beam thrust, transverse thrust, spherocity, 
and F-parameter, which are in particular sensitive to properties of the underlying event at small values of 
the Z-boson transverse momentum. 
A detector-level measurement of track distributions sensitive to the properties of the underlying event is presented in Ref.\ \cite{Und_Tr}. 
It is based on an integrated luminosity of 151~$\mu$b$^{-1}$
of pp-collisions at a centre-of-mass energy of $\sqrt{s} = 13$~TeV. 
Distributions of the track multiplicity and of the track transverse momentum are measured in regions of azimuthal 
angle defined with respect to the leading track. 
The measured distributions are compared with predictions from Monte Carlo event generators.
Earlier the underlying events for  $Z$-boson and jets 
at 7~TeV were studied with ATLAS detector (see Ref.\ \cite{Kulchitsky:2016mfm}).

\vspace*{-5mm}
%_______________________________________________________________________
\section{Event-shape observables for Drell-Yan events
}
\label{sec-2}
%_______________________________________________________________________
\vspace*{-2mm}

Events were selected by requiring a Z-boson candidate decaying to an $e^+e^-$  or $\mu^+\mu^-$ pair.
The charged-particle event-shape observables beam thrust, transverse thrust, spherocity, and
$F$-parameter as defined below were measured in inclusive Z production. 
The observables were calculated for primary charged particles with  $p_{\rm T} > 0.5$~GeV  and  $|\eta| < 2.5$. 
At small $p_{\rm T} (l^+ l^-)$ values, events are expected to have low jet activity from the hard process and hence high sensitivity to
UE characteristics. 
At high $p_{\rm T} (l^+ l^-)$ values, the event is expected to contain at least one jet of high transverse momentum 
recoiling against the $l^+ l^-$ system, which is expected to be reasonably described by perturbative calculations of the
hard process.

Distributions $f_{O} = 1/N_{ev} \cdot dN/dO$ were measured for all selected events, $N_{ev}$, for the following observables $O$.
1) The charged-particle multiplicity, $N_{ch}$.
2) The scalar sum of transverse momenta of selected charged particles, $\sum_i p_{\rm T,  i} = \sum p_{\rm T}$.
3) The beam thrust, $B$, as proposed in Refs.\ \cite{UE_shape_20,UE_shape_21,UE_shape_22}. 
This is similar to $\sum p_{\rm T}$ except that in the sum over all charged particles the transverse momentum of each particle is
weighted by a factor depending on its  $\eta$:
$B = \sum_i p_{\rm T,  i} \cdot e^{-|\eta_i|}$.
As a result, contributions from particles in the forward and backward direction 
are suppressed with respect to particles emitted at central $\eta$.
4) The transverse thrust, $T$, as proposed in Ref.\  \cite{UE_shape_23}:
$T = \max_{\vec{n}_{\rm T}} \frac{\sum_i | \vec{p}_{\rm T,  i} \cdot \vec{n}_{\rm T} |}{\sum_i p_{\rm T,  i} }  $,
where the sum runs over all charged particles, and the thrust axis, $\vec{n}_{\rm T}$, maximises the expression. 
The solution for $\vec{n}_{\rm T}$ is found iteratively following the algorithm described in Ref.\ \cite{UE_shape_24}.
5) The spherocity, $S$, as proposed in Ref.\  \cite{UE_shape_23}:
$S = \frac{\pi^2}{4} \min_{\vec{n}=(n_x, n_y, 0)^{\rm T}} 
\left( \frac{ \sum_i \left|\vec{p}_{\rm T, i} \times \vec{n} \right|}{\sum_i \vec{p}_{\rm T, i}}\right)^2  $,
where the sum runs over all charged particles and the vector $\vec{n}$ minimises the expression. 
6) The $F$-parameter defined as the ratio of the smaller and larger eigenvalues, $\lambda_1$ and $\lambda_2$
$F= \frac{\lambda_1}{\lambda_2}$
of the transverse momentum tensor $M^{lin}$ in Ref.\ \cite{Aad:2016ria}. 
Pencil-like events, e.g.\ containing two partons emitted in opposite directions in the transverse plane, are characterised
by values of $S$, $T$, and $F$ close to 0, 1, and 0, respectively.
The corresponding values of these observables for spherical events, e.g.\ containing several partons emitted isotropically,
are close to 1, $2/\pi$, and 1, respectively. 
While the event-shape observables $S$, $T$, and $F$ show very high correlations among
themselves, they are weakly correlated with $N_{ch}$, $\vec{p}_{\rm T}$, and $B$.
Each observable was determined in the following ranges of the transverse momentum of the Z boson, $p_{\rm T} (l^+l^−)$, 
calculated from the four-momenta of the 
$l^+l^−$ pair: 0--6, 6--12, 1--25, and $\ge 25$~GeV. 
Events at small $p_{\rm T} (l^+l^−)$ are expected to be particularly sensitive to the UE activity, while events with large 
$p_{\rm T} (l^+l^−)$ values 
are expected to contain significant contributions from jet production 
coming from the hard-scattering process. 
%

%_______________________________________________________________________
%fig 1
\begin{figure*}[t!]
\begin{minipage}[h]{0.3\textwidth}
\center{\includegraphics[width=1.0\linewidth,height=0.2\textheight]{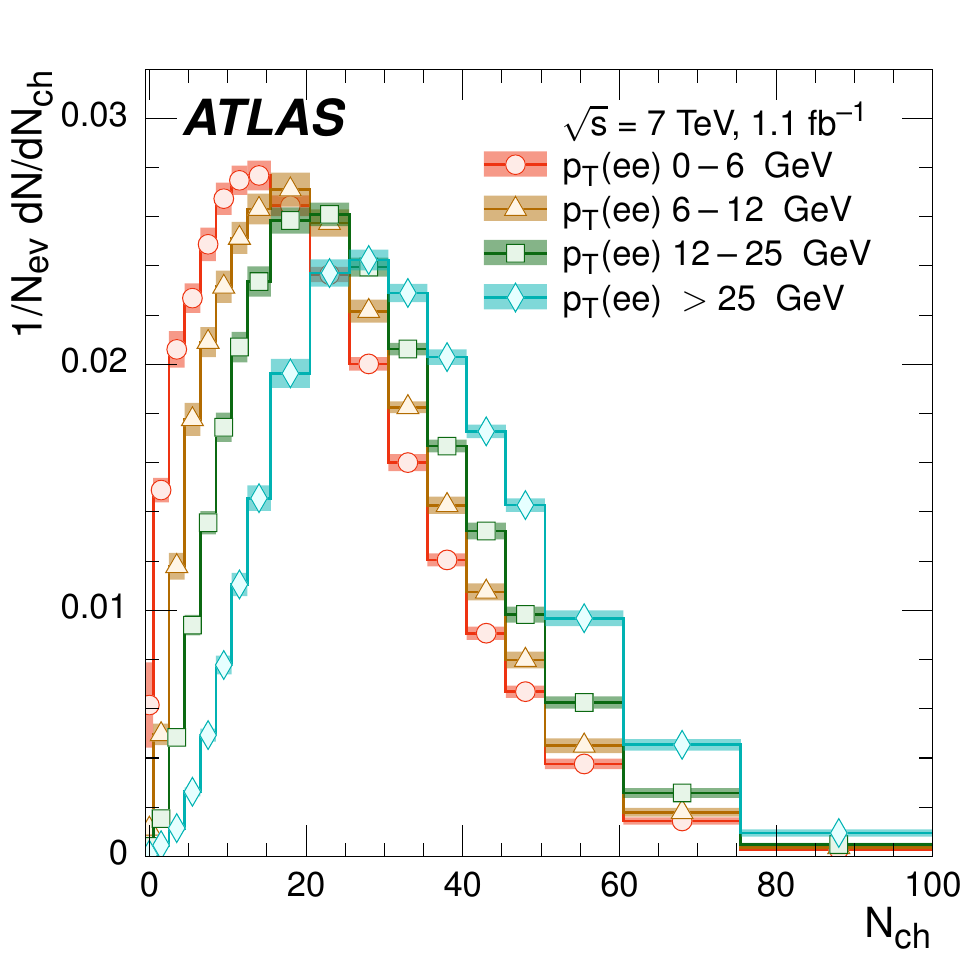}}
(a) \\
\end{minipage}
\hfill
\begin{minipage}[h]{0.3\textwidth} 
\center{\includegraphics[width=1.0\linewidth,height=0.2\textheight]{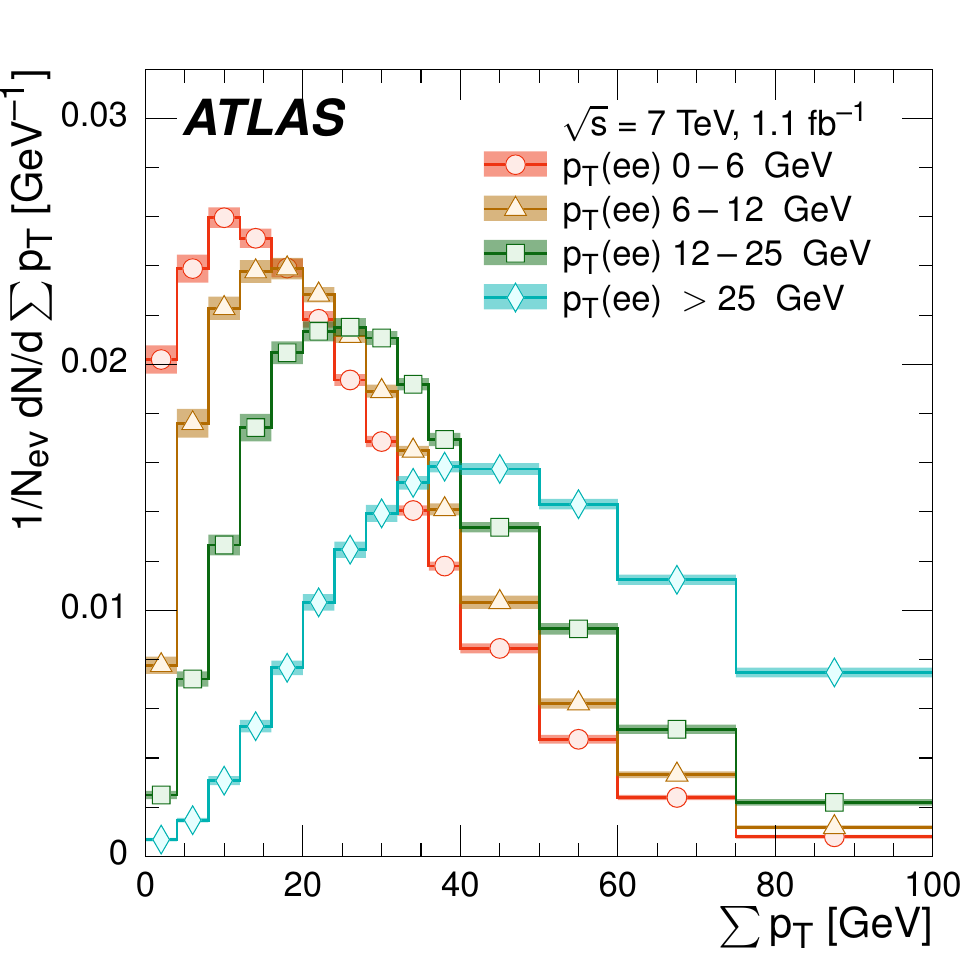}}
(b) \\
\end{minipage}
\hfill
\begin{minipage}[h]{0.3\textwidth}
\center{\includegraphics[width=1\linewidth,height=0.2\textheight]{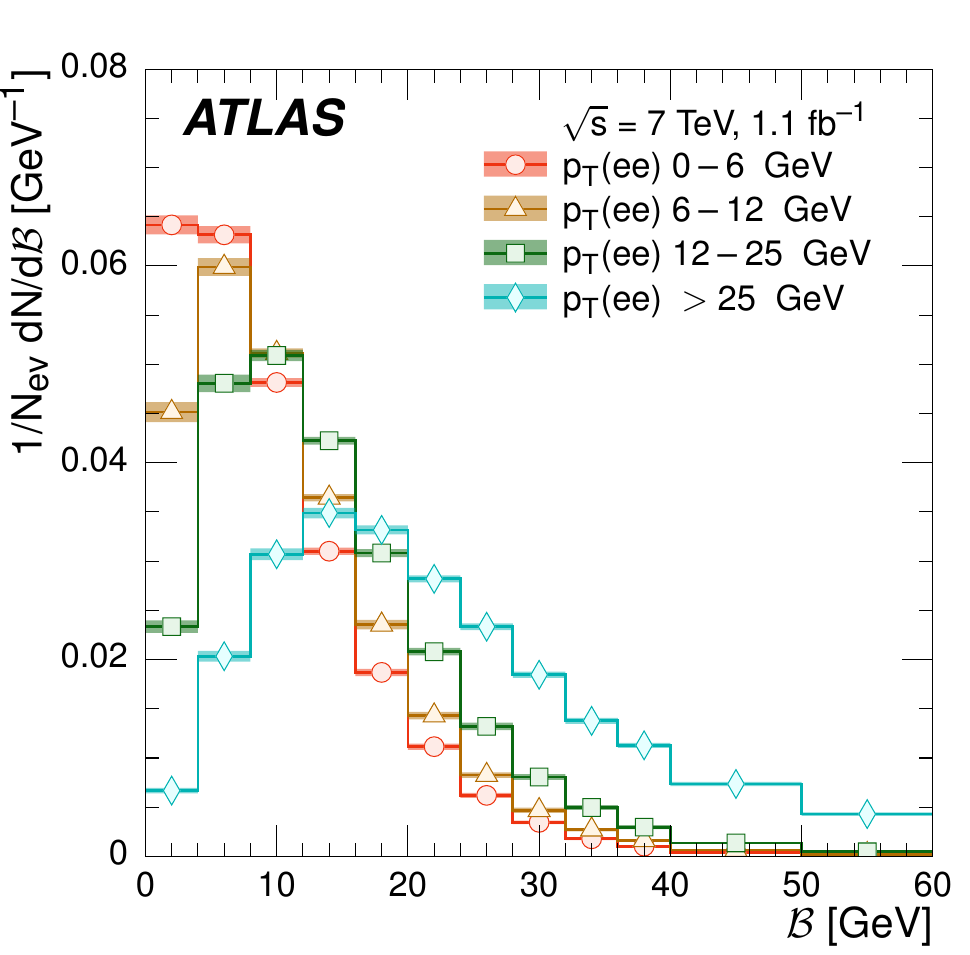}} 
(c) \\
\end{minipage}
\vfill
\begin{minipage}[h]{0.3\textwidth} 
\center{\includegraphics[width=1\linewidth,height=0.2\textheight]{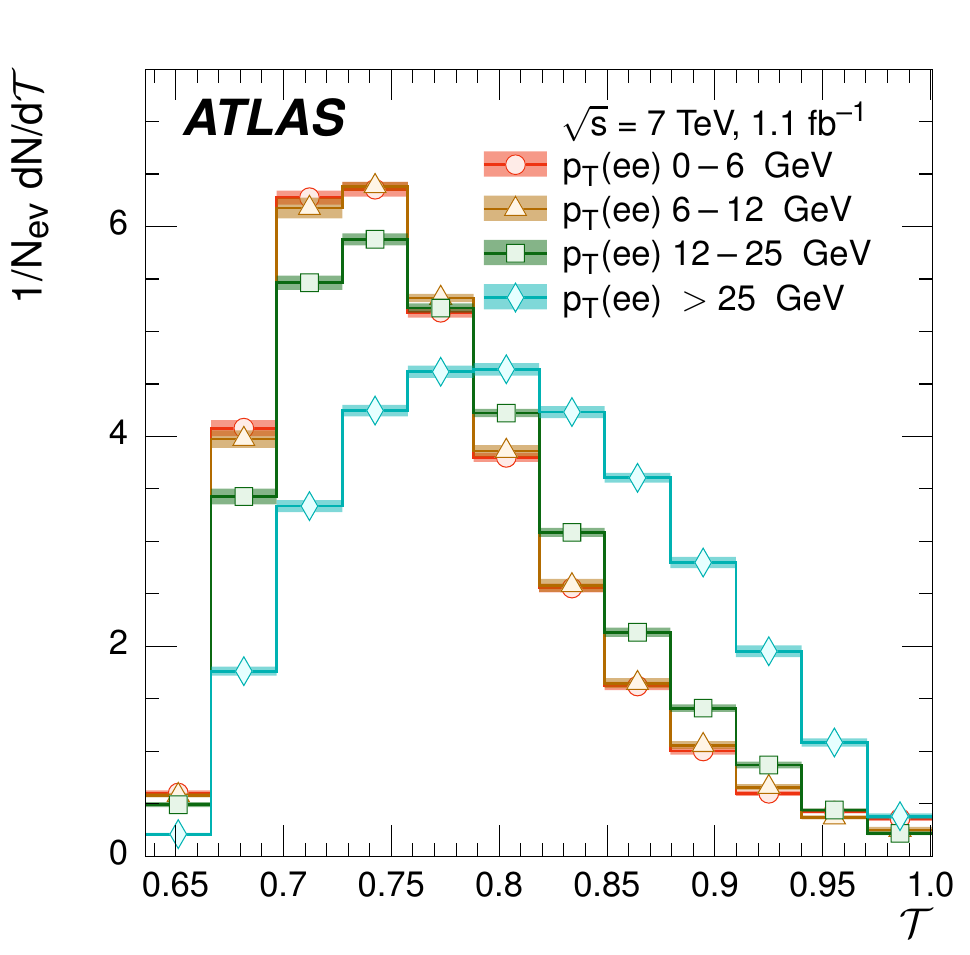}} 
(d) \\
\end{minipage}
\hfill
\begin{minipage}[h]{0.3\textwidth}
\center{\includegraphics[width=1.0\linewidth,height=0.2\textheight]{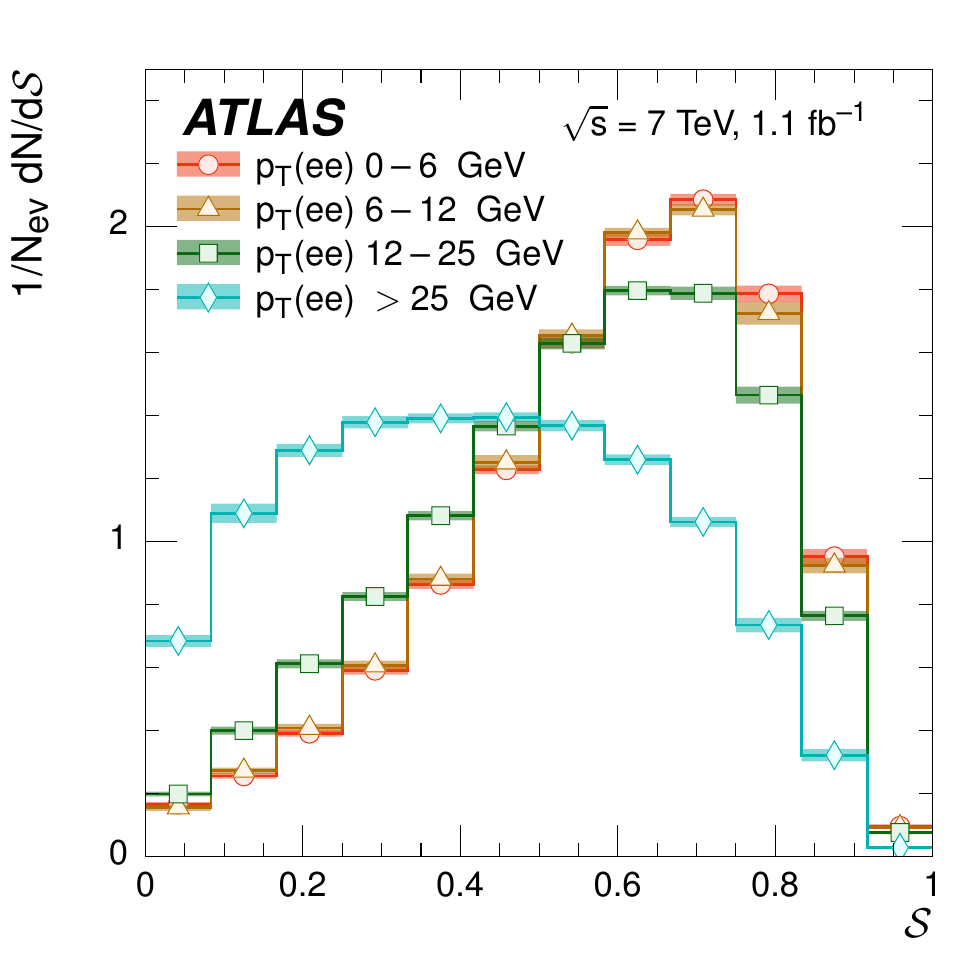}}
(e) \\
\end{minipage}
\hfill
\begin{minipage}[h]{0.3\textwidth} 
\center{\includegraphics[width=1.0\linewidth,height=0.2\textheight]{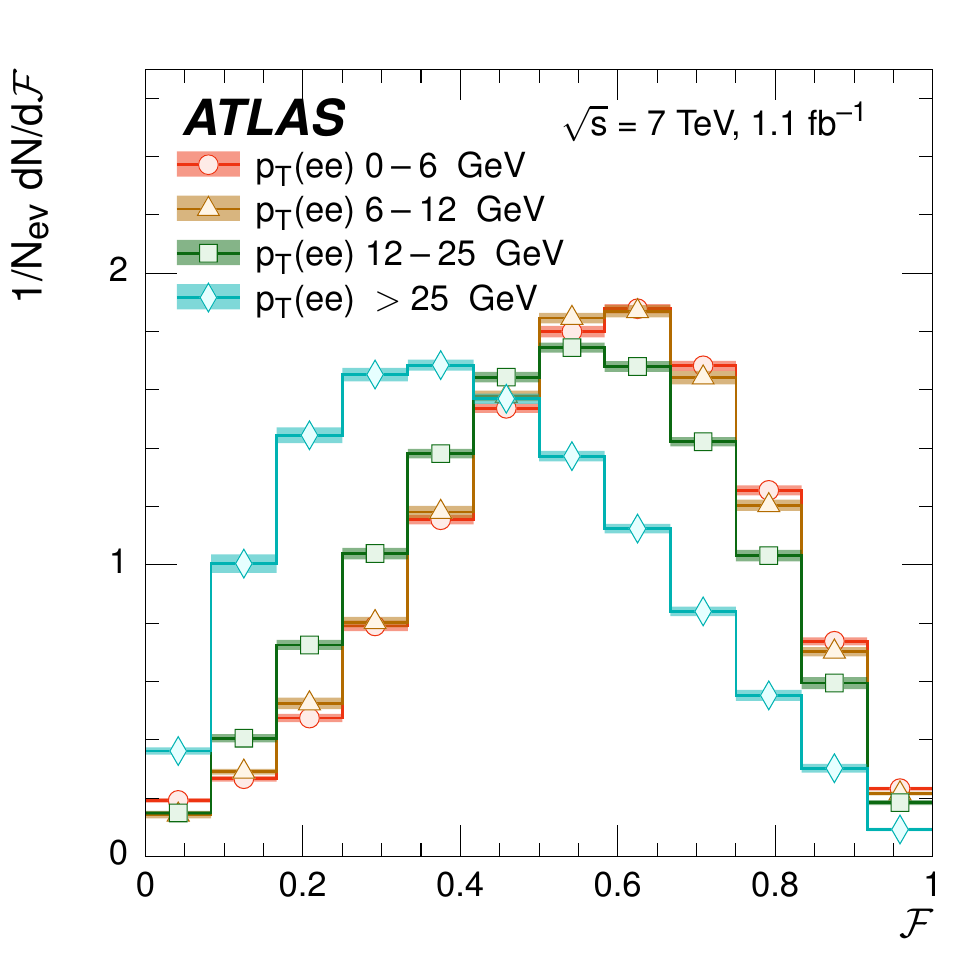}}
(f) \\
\end{minipage}
\caption{
Distributions of the event-shape variables 
(a) charged-particle multiplicity, 
(b) summed transverse momenta, 
(c) beam thrust, 
(d) transverse thrust, 
(e) spherocity, 
and 
(f) $F$-parameter 
measured in $Z\rightarrow e^+ e^-$ events for the different ranges of 
$p_{\rm T} (e^+ e^-)$.
Taken from Ref.\ \cite{Aad:2016ria}.
}
\label{fig_Zll_1}
\end{figure*}
%_______________________________________________________________________

%
For comparison with corrected distributions, three 
recent versions of MC event  generators were used 
to provide predictions for the signal at particle level: 
%1) 
Sherpa 2.2.0  with up to two additional partons at NLO and with three additional partons at LO and 
taking the NLO matrix element calculations for virtual contributions from OpenLoops \cite{UE_shape_36}
with the NNPDF 3.0 NNLO PDF set \cite{UE_shape_37}; 
%2) 
Pythia~8.212 with LO matrix element calculations using the NNPDF2.3 LO PDF set \cite{UE_shape_38}; 
%3) 
Herwig~7.0 \cite{UE_shape_1}   taking the NLO matrix element calculations for real emissions from MadGraph \cite{UE_shape_39} 
and for virtual contributions from OpenLoops using the MMHT2014  PDF set \cite{UE_shape_40}. 

%_______________________________________________________________________
%fig 2
\begin{figure*}[t!]
\begin{minipage}[h]{0.24\textwidth}
\center{\includegraphics[width=1.0\linewidth,height=0.2\textheight]{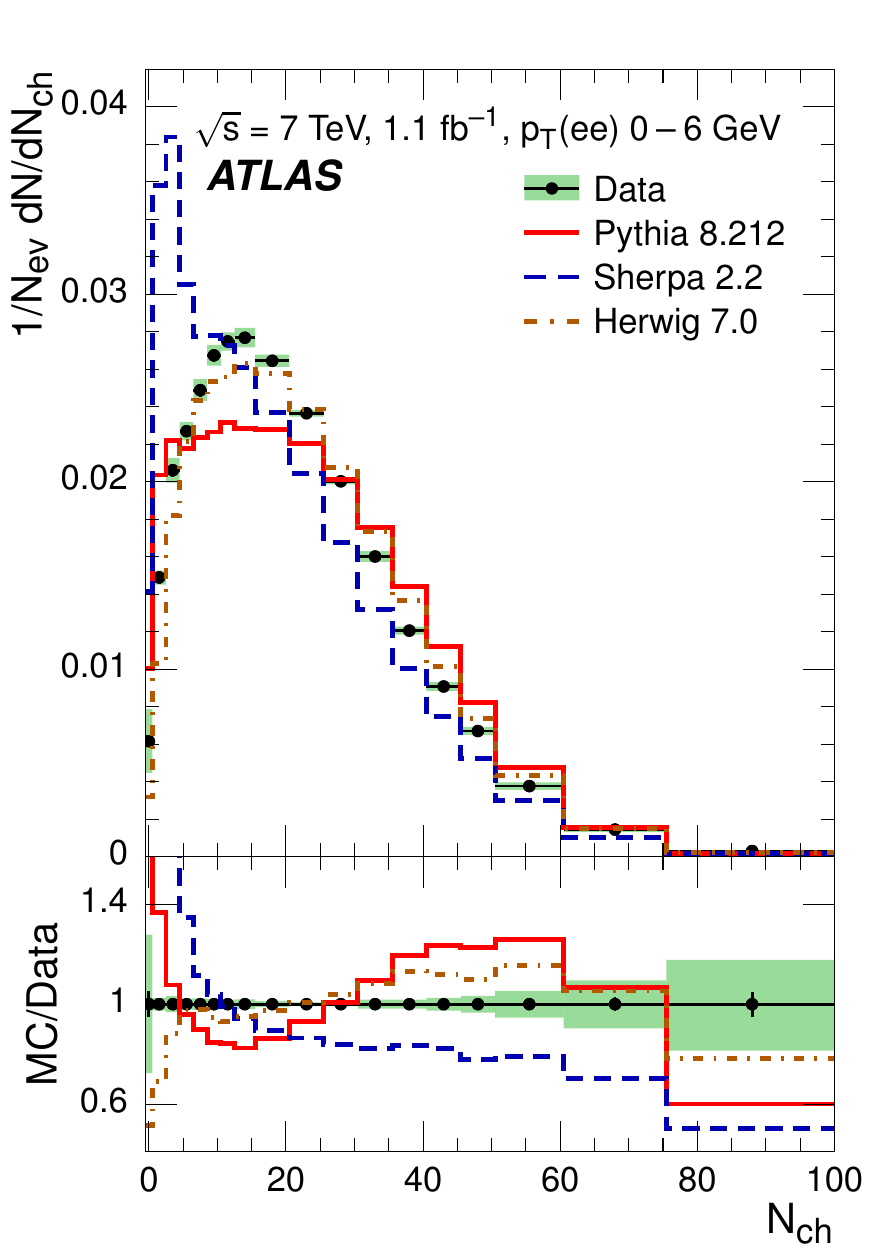}}
(a) \\
\end{minipage}
\hfill
\begin{minipage}[h]{0.24\textwidth} 
\center{\includegraphics[width=1.0\linewidth,height=0.2\textheight]{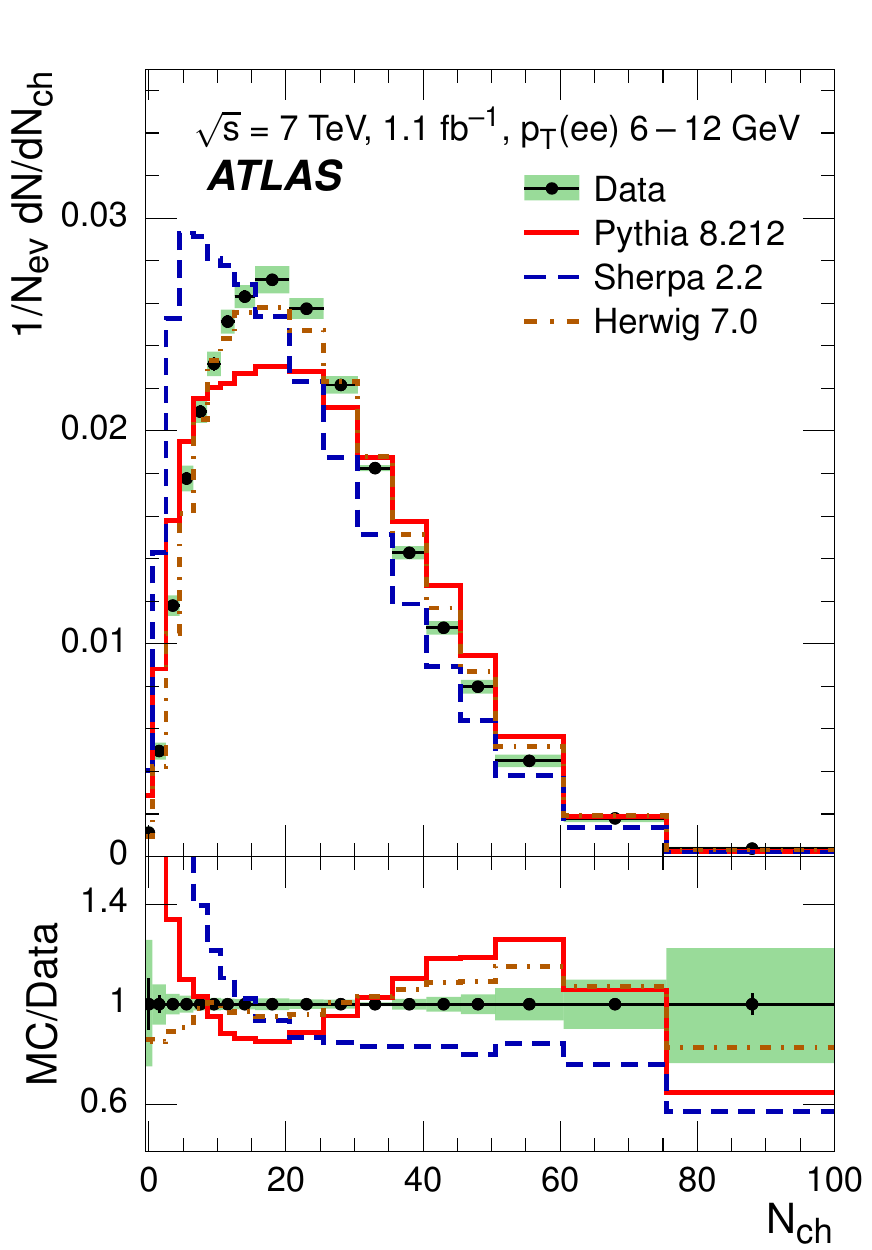}}
(b) \\
\end{minipage}
\hfill
\begin{minipage}[h]{0.24\textwidth}
\center{\includegraphics[width=1\linewidth,height=0.2\textheight]{fig_03b.pdf}} 
(c) \\
\end{minipage}
\hfill
\begin{minipage}[h]{0.24\textwidth} 
\center{\includegraphics[width=1\linewidth,height=0.2\textheight]{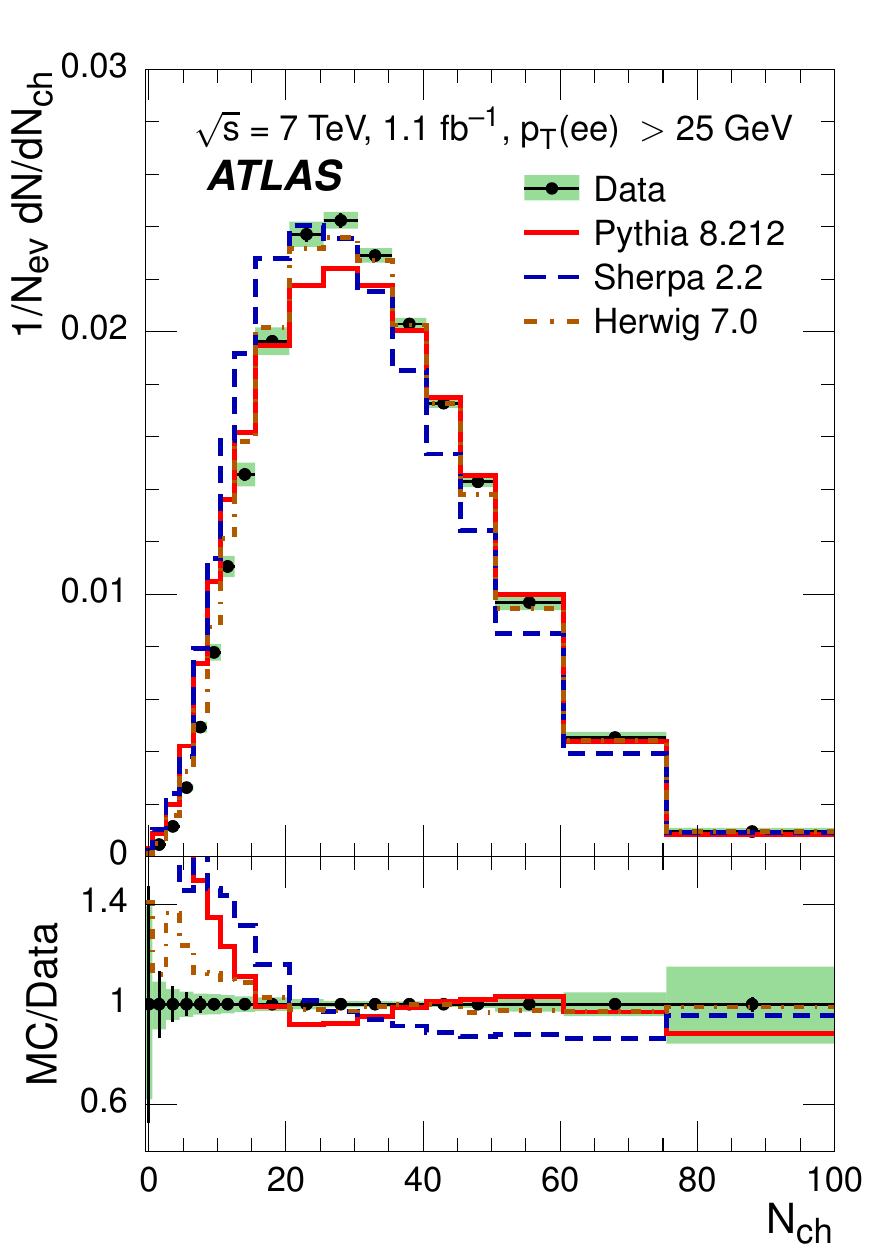}} 
(d) \\
\end{minipage}
\vfill
\begin{minipage}[h]{0.24\textwidth}
\center{\includegraphics[width=1.0\linewidth,height=0.2\textheight]{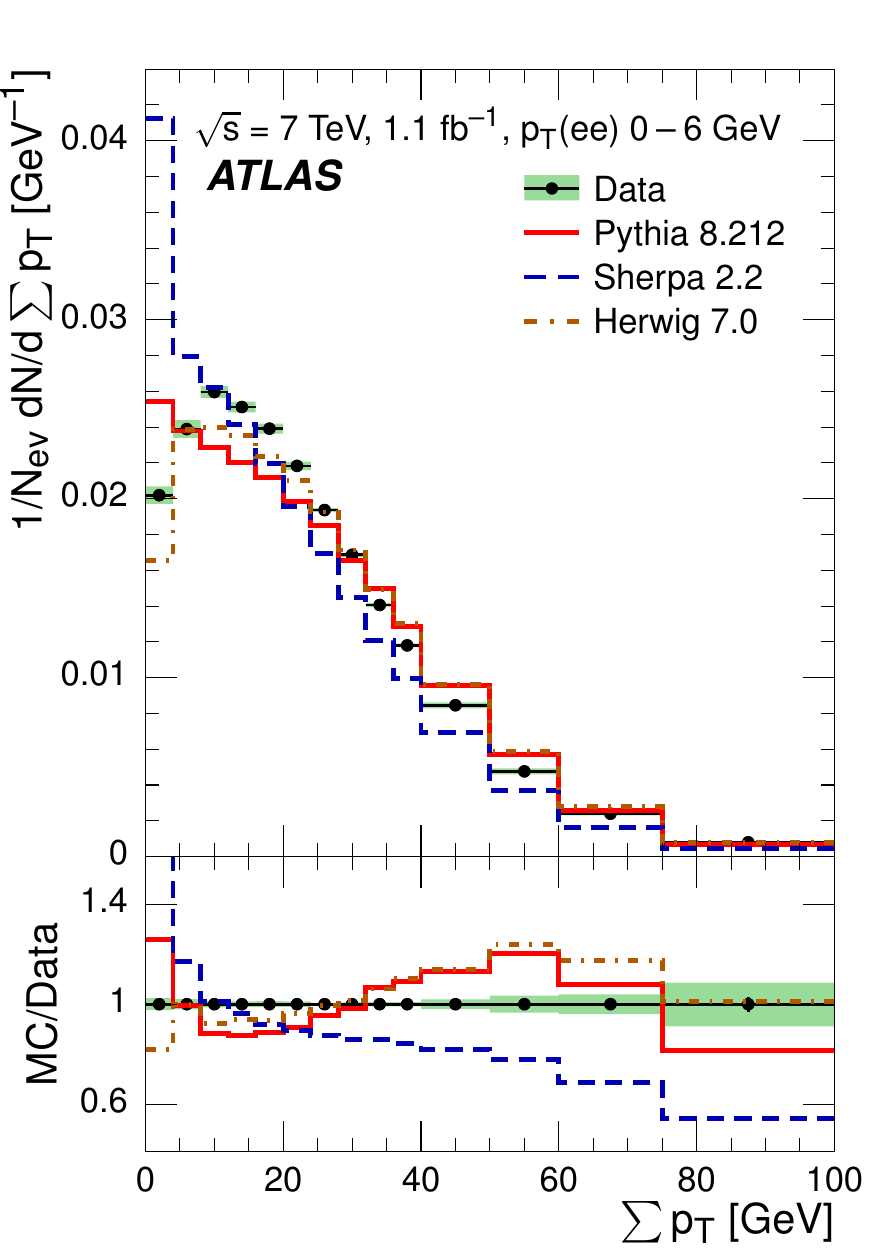}}
(e) \\
\end{minipage}
\hfill
\begin{minipage}[h]{0.24\textwidth} 
\center{\includegraphics[width=1.0\linewidth,height=0.2\textheight]{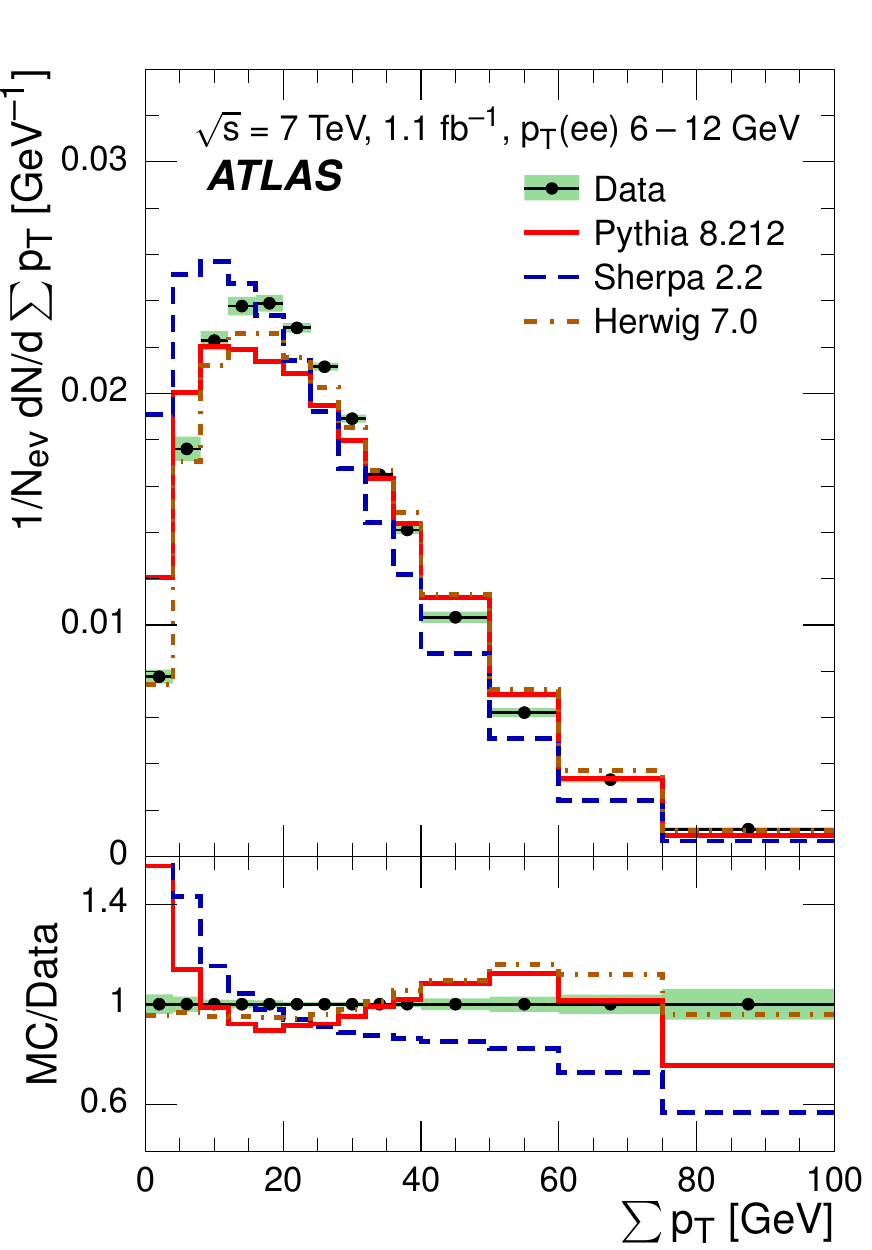}}
(f) \\
\end{minipage}
\hfill
\begin{minipage}[h]{0.24\textwidth}
\center{\includegraphics[width=1\linewidth,height=0.2\textheight]{fig_04b.pdf}} 
(g) \\
\end{minipage}
\hfill
\begin{minipage}[h]{0.24\textwidth} 
\center{\includegraphics[width=1\linewidth,height=0.2\textheight]{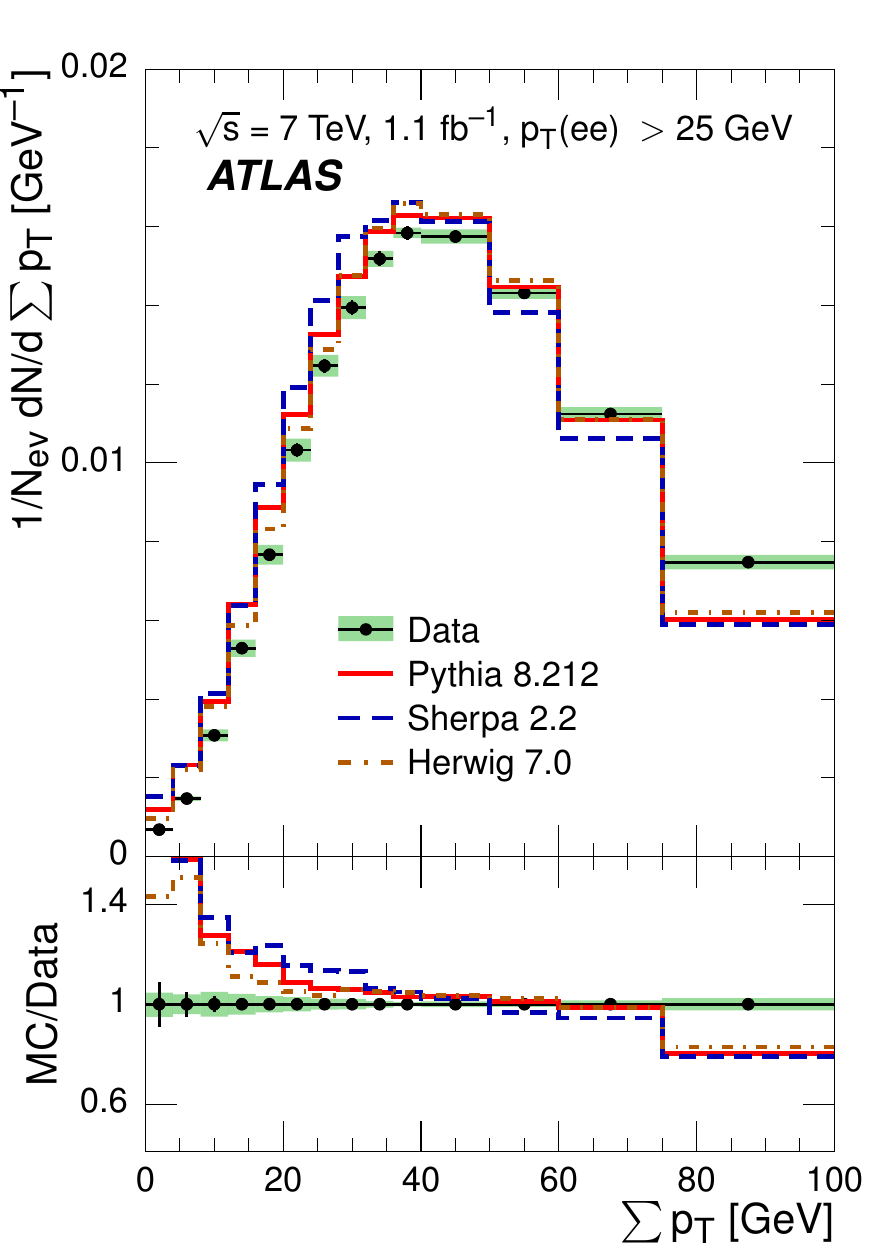}} 
(h) \\
\end{minipage}
\vfill
\begin{minipage}[h]{0.24\textwidth}
\center{\includegraphics[width=1.0\linewidth,height=0.2\textheight]{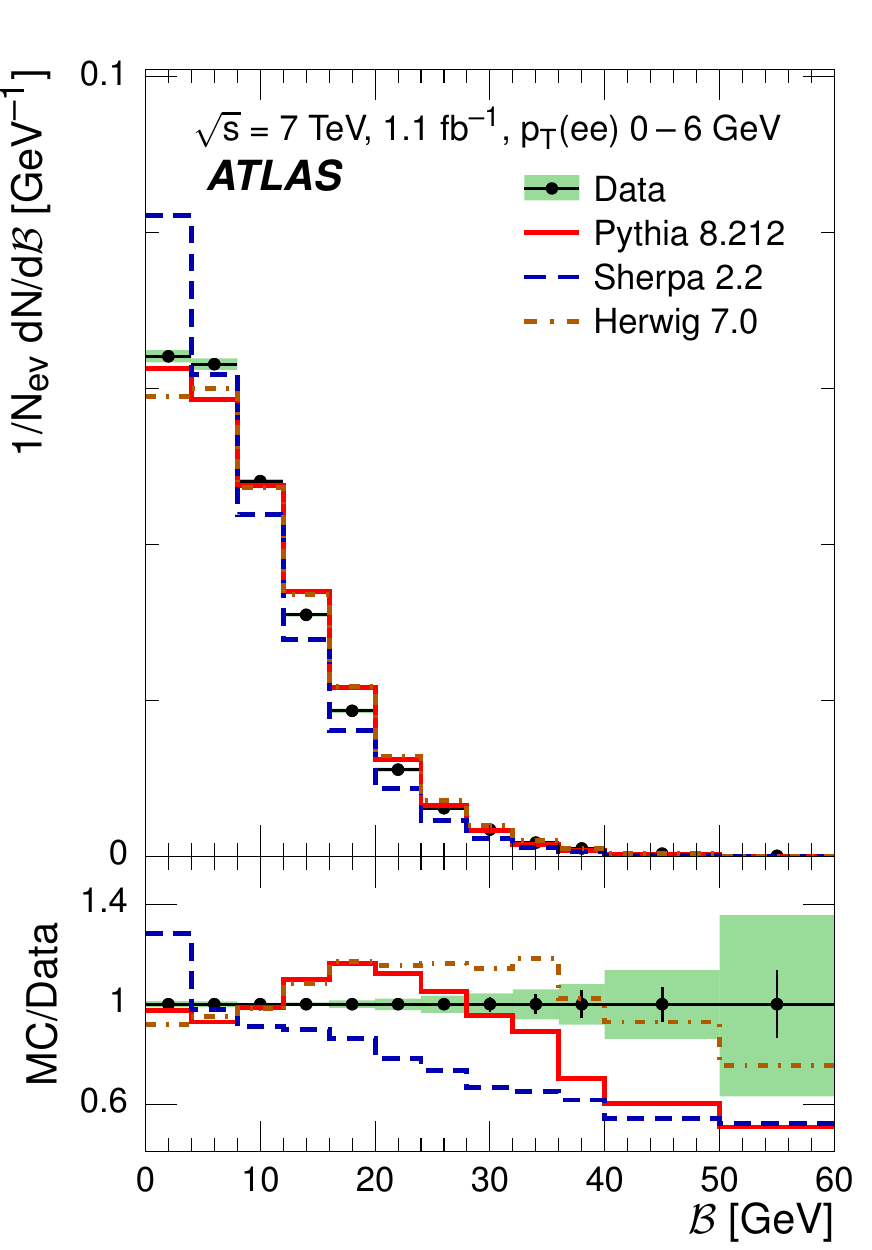}}
(i) \\
\end{minipage}
\hfill
\begin{minipage}[h]{0.24\textwidth} 
\center{\includegraphics[width=1.0\linewidth,height=0.2\textheight]{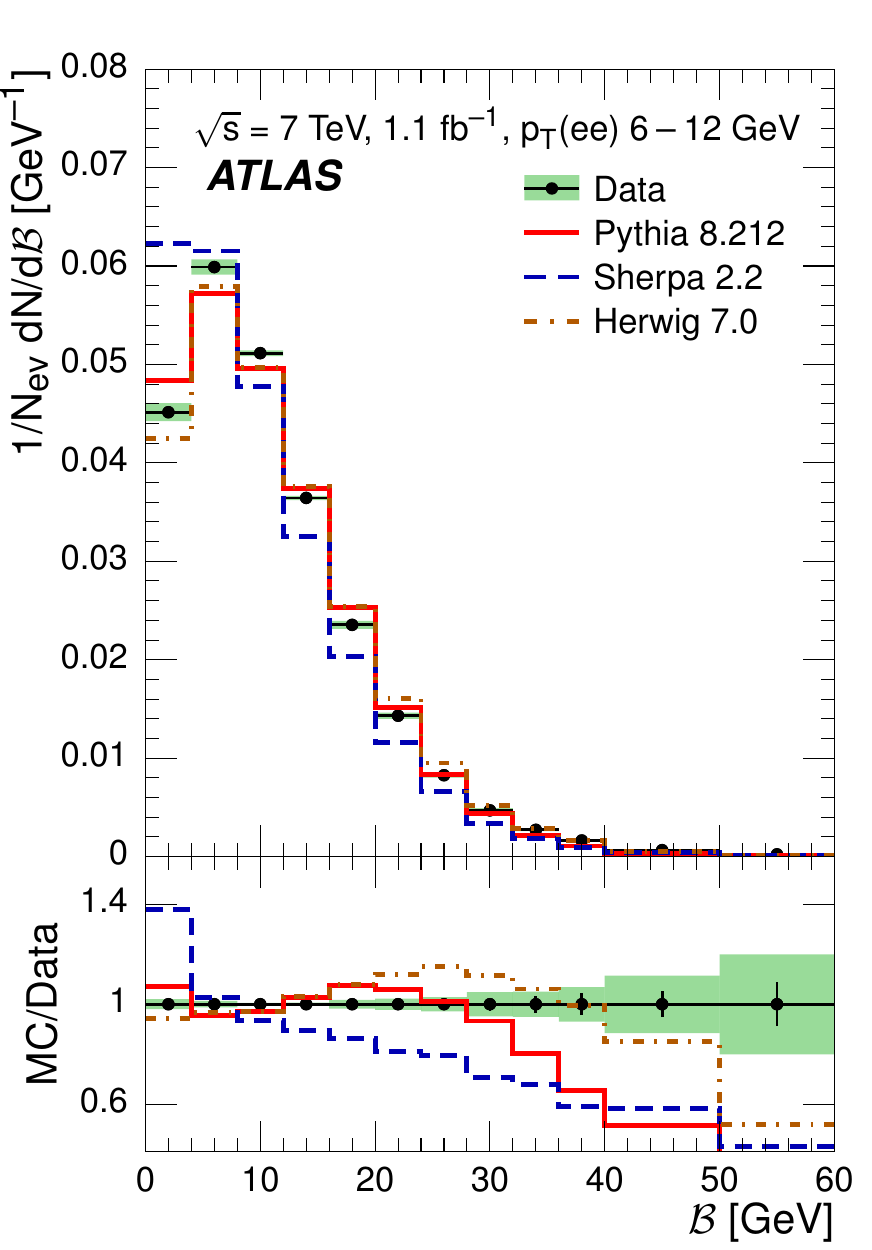}}
(j) \\
\end{minipage}
\hfill
\begin{minipage}[h]{0.24\textwidth}
\center{\includegraphics[width=1\linewidth,height=0.2\textheight]{fig_05b.pdf}} 
(k) \\
\end{minipage}
\hfill
\begin{minipage}[h]{0.24\textwidth} 
\center{\includegraphics[width=1\linewidth,height=0.2\textheight]{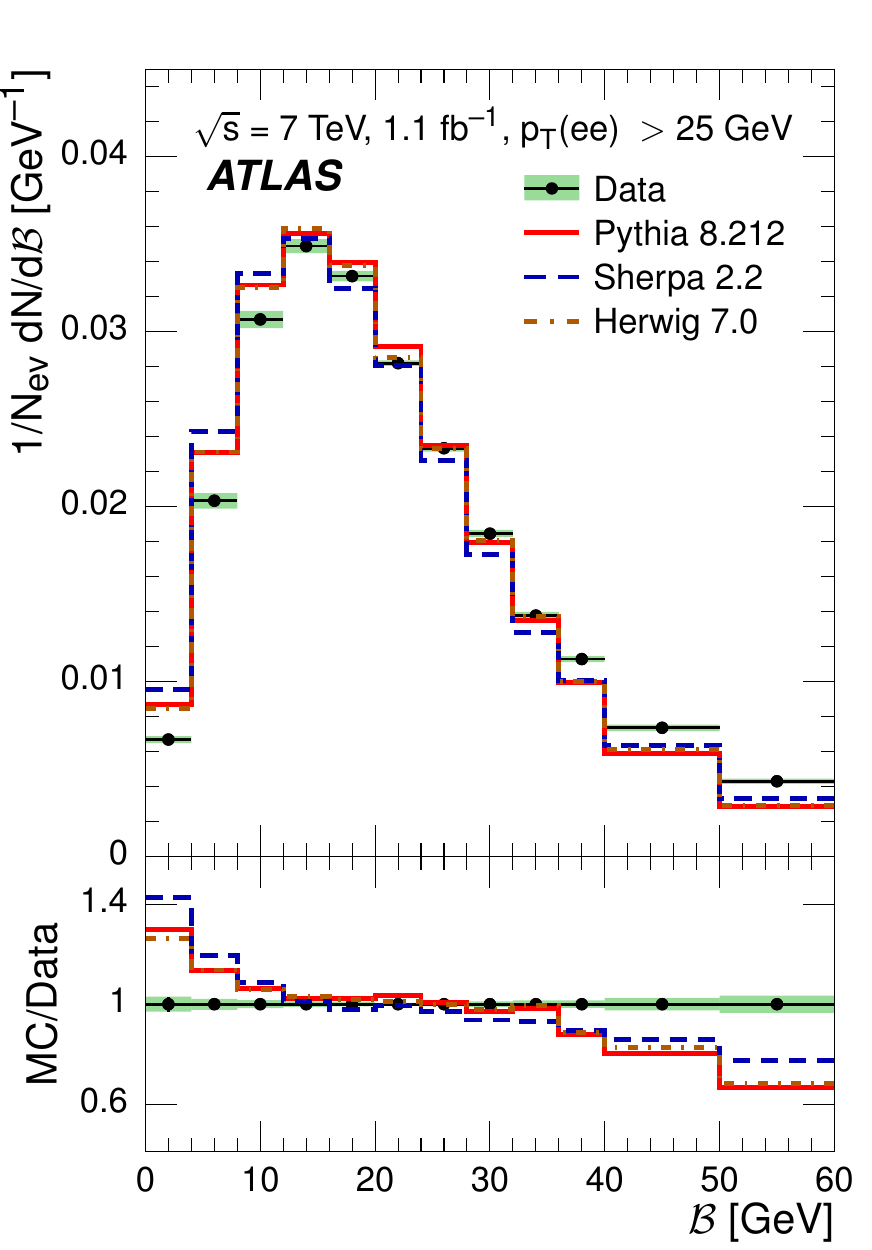}} 
(l) \\
\end{minipage}
\caption{
(a--d) Multiplicity,
(e--h) summed transverse momenta 
and
(i--l) beam thrust 
distributions of charged particles  for $Z\rightarrow e^+ e^-$ 
for the  $p_{\rm T} (e^+ e^-)$  ranges 
compared to the predictions from the MC generators.
Taken from Ref.\ \cite{Aad:2016ria}.
}
\label{fig_Zll_2}
\end{figure*}
%_______________________________________________________________________
%
%_______________________________________________________________________
%fig 3
\begin{figure*}[t!]
\begin{minipage}[h]{0.24\textwidth}
\center{\includegraphics[width=1.0\linewidth,height=0.2\textheight]{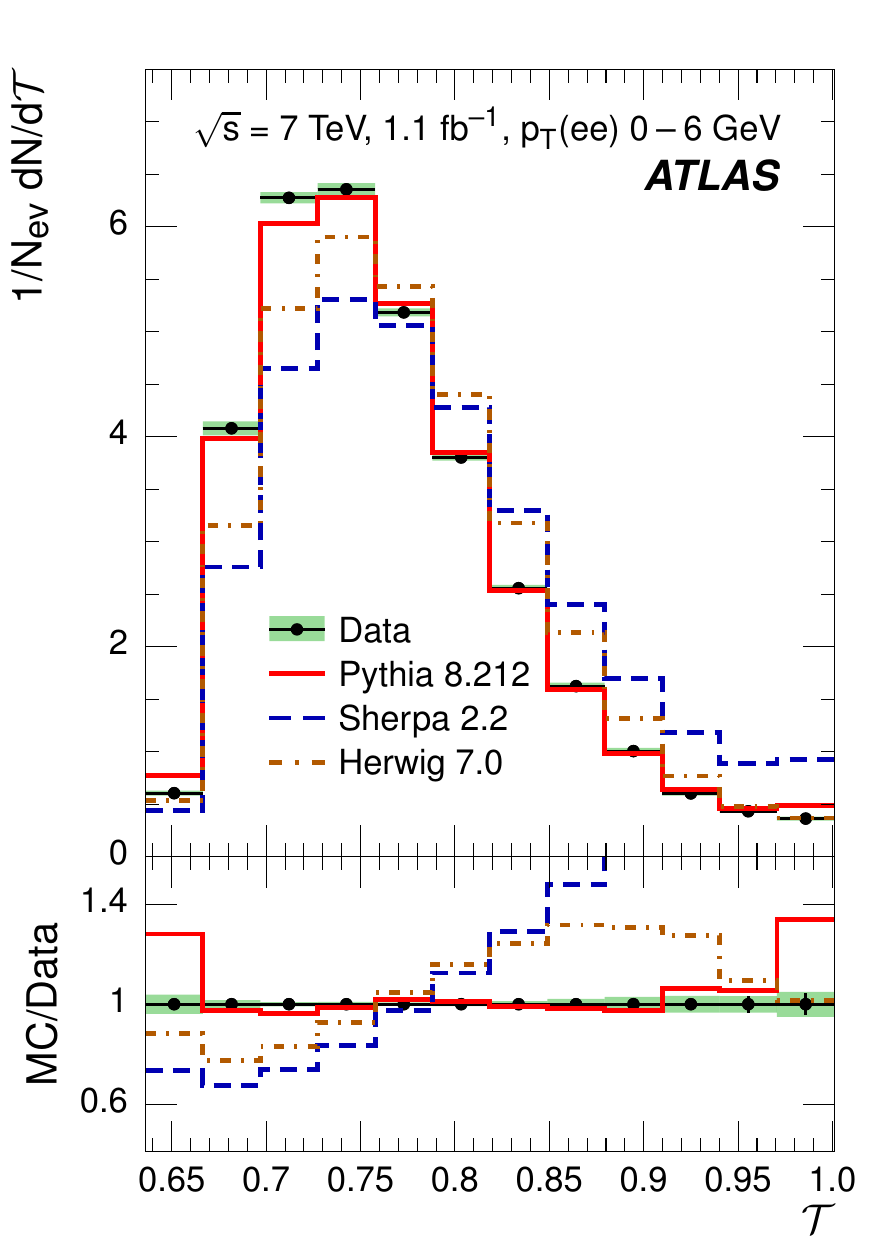}}
(a) \\
\end{minipage}
\hfill
\begin{minipage}[h]{0.24\textwidth} 
\center{\includegraphics[width=1.0\linewidth,height=0.2\textheight]{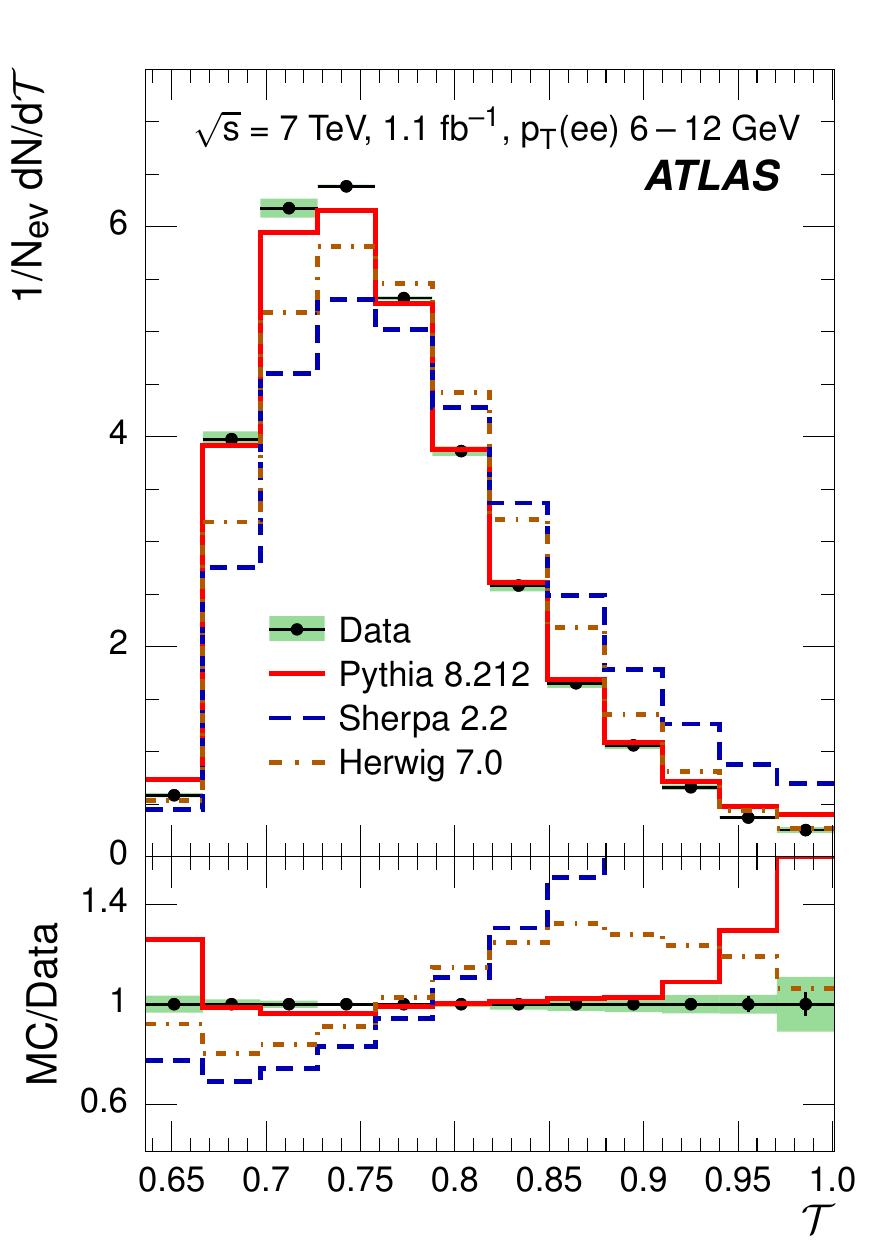}}
(b) \\
\end{minipage}
\hfill
\begin{minipage}[h]{0.24\textwidth}
\center{\includegraphics[width=1\linewidth,height=0.2\textheight]{fig_06b.pdf}} 
(c) \\
\end{minipage}
\hfill
\begin{minipage}[h]{0.24\textwidth} 
\center{\includegraphics[width=1\linewidth,height=0.2\textheight]{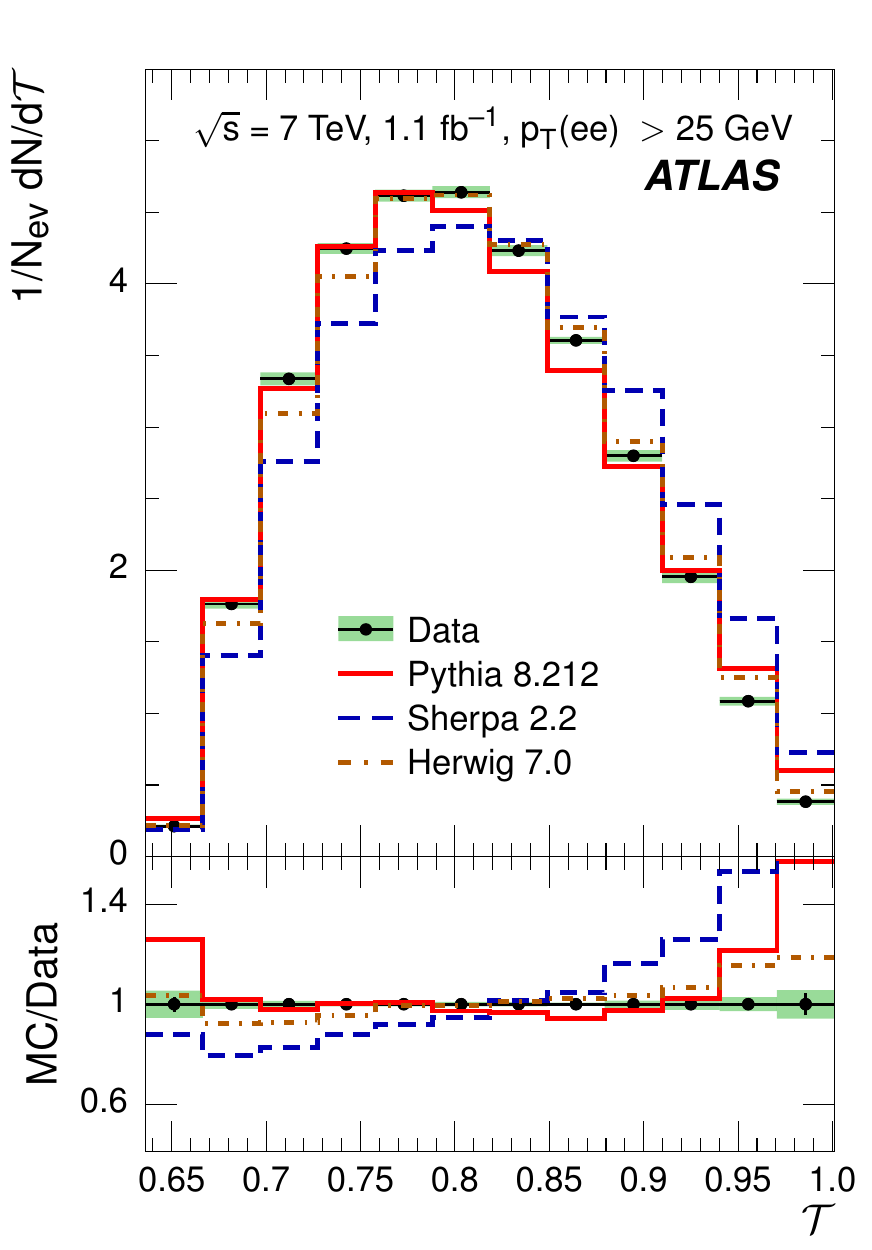}} 
(d) \\
\end{minipage}
\vfill
\begin{minipage}[h]{0.24\textwidth}
\center{\includegraphics[width=1.0\linewidth,height=0.2\textheight]{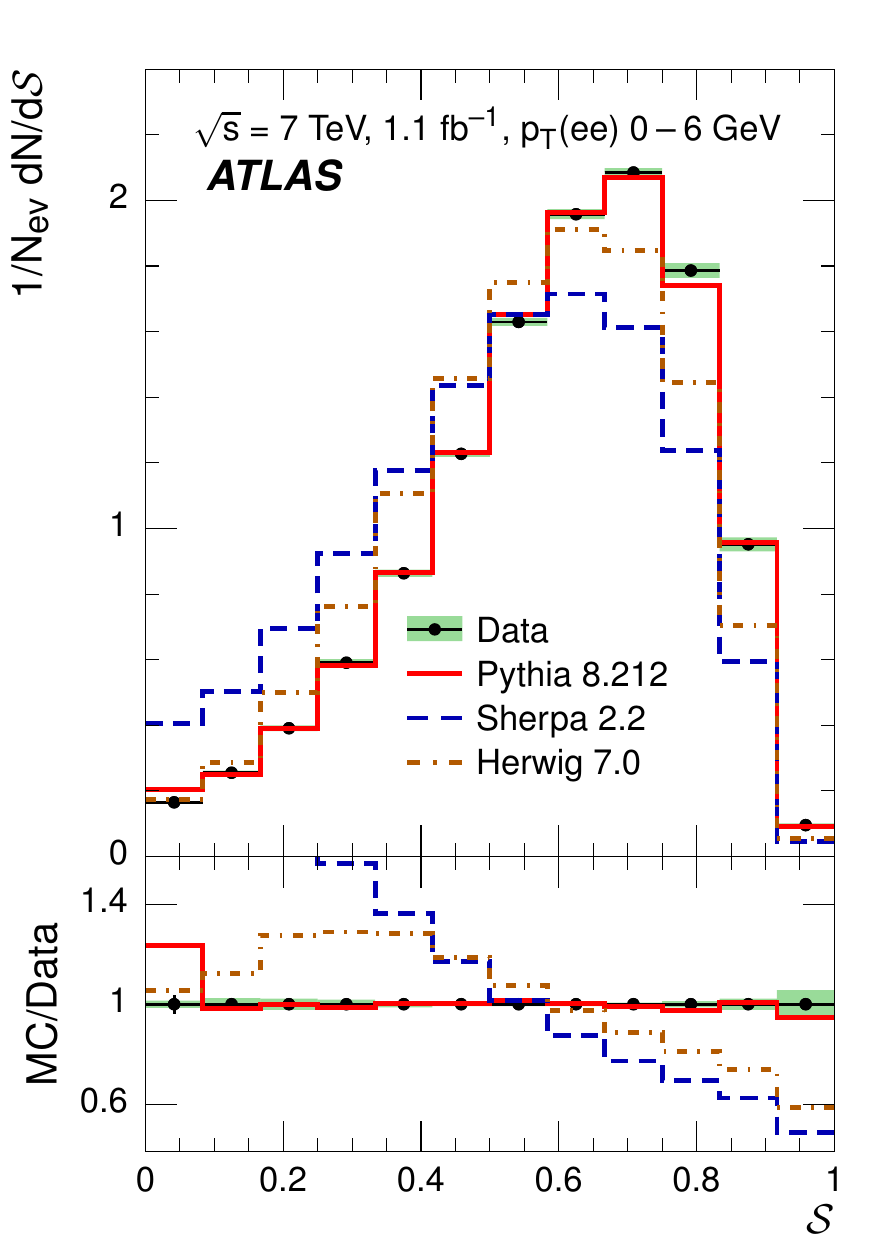}}
(e) \\
\end{minipage}
\hfill
\begin{minipage}[h]{0.24\textwidth} 
\center{\includegraphics[width=1.0\linewidth,height=0.2\textheight]{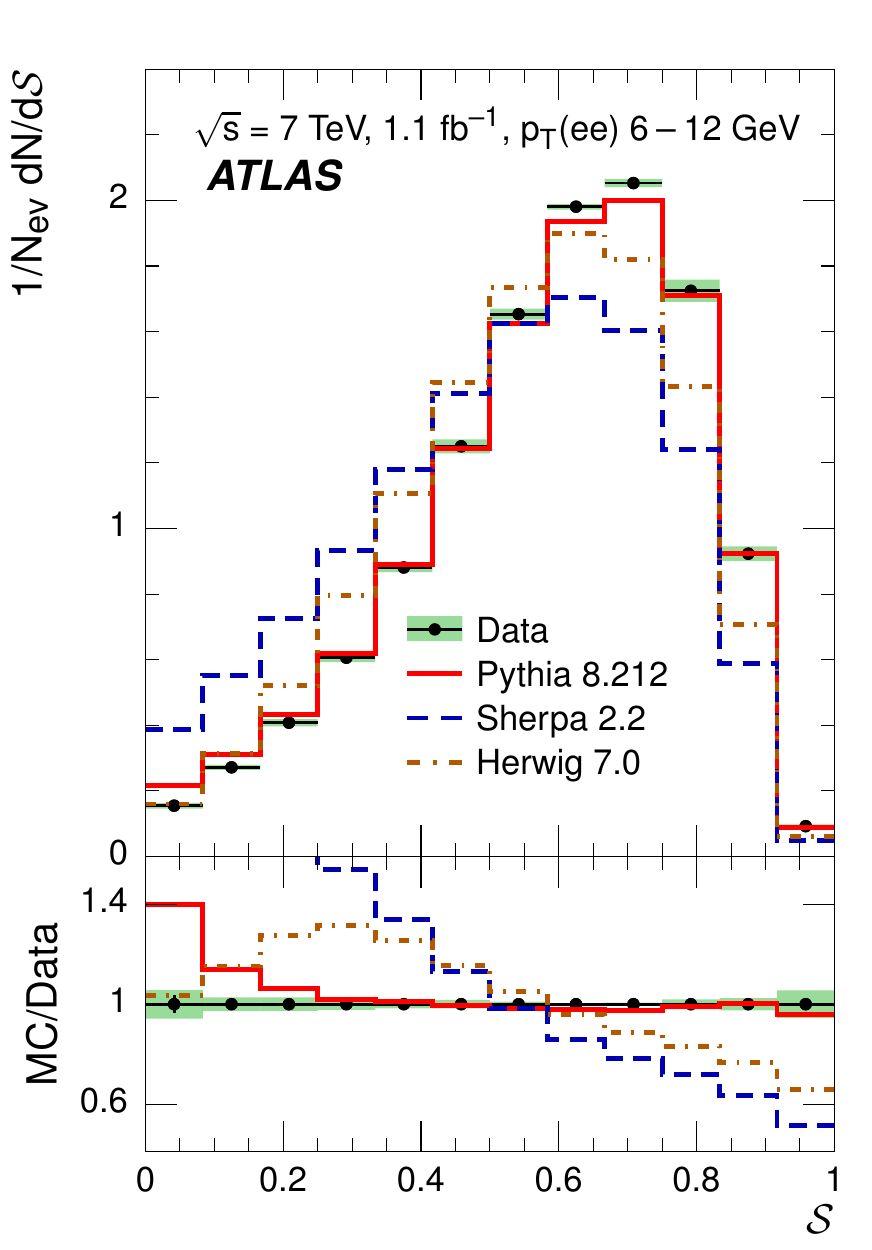}}
(f) \\
\end{minipage}
\hfill
\begin{minipage}[h]{0.24\textwidth}
\center{\includegraphics[width=1\linewidth,height=0.2\textheight]{fig_07b.pdf}} 
(g) \\
\end{minipage}
\hfill
\begin{minipage}[h]{0.24\textwidth} 
\center{\includegraphics[width=1\linewidth,height=0.2\textheight]{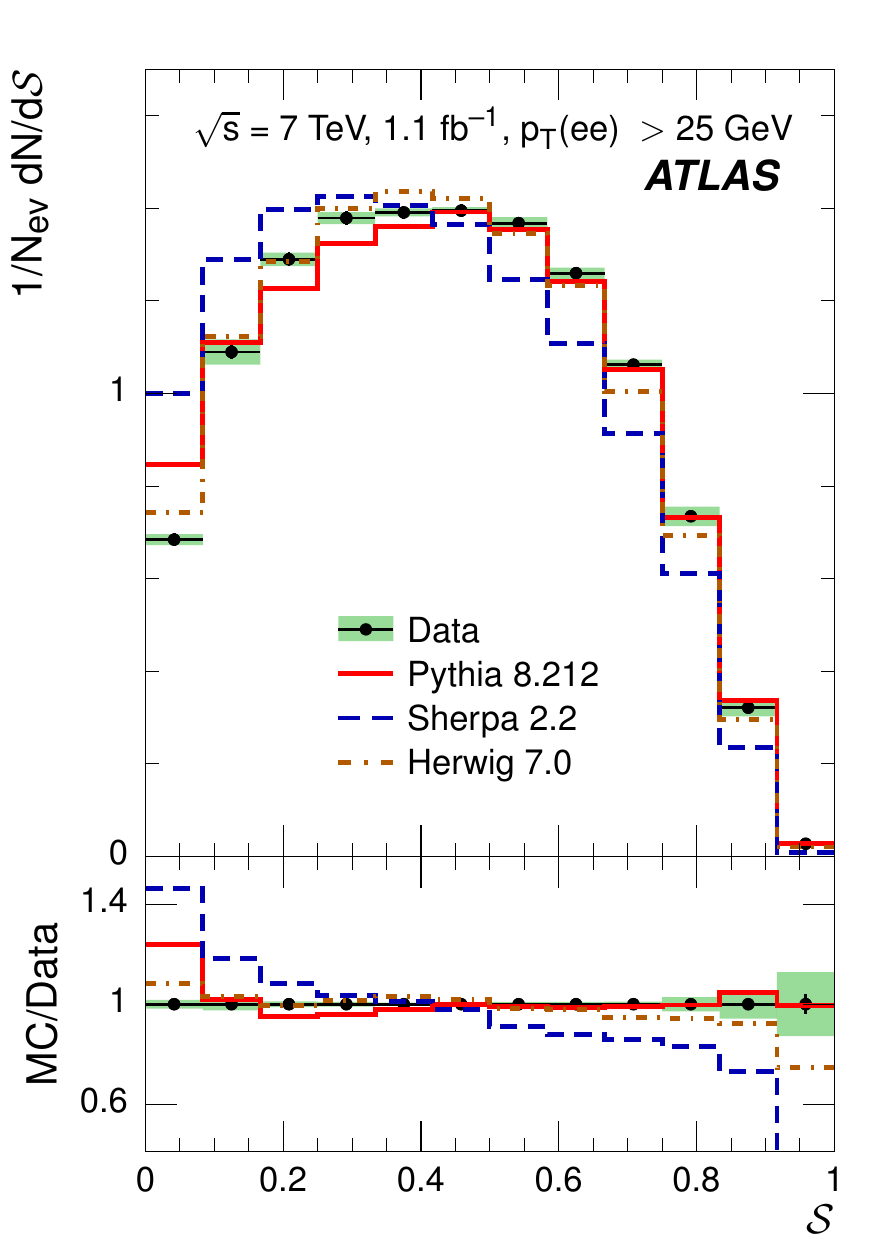}} 
(h) \\
\end{minipage}
\vfill
\begin{minipage}[h]{0.24\textwidth}
\center{\includegraphics[width=1.0\linewidth,height=0.2\textheight]{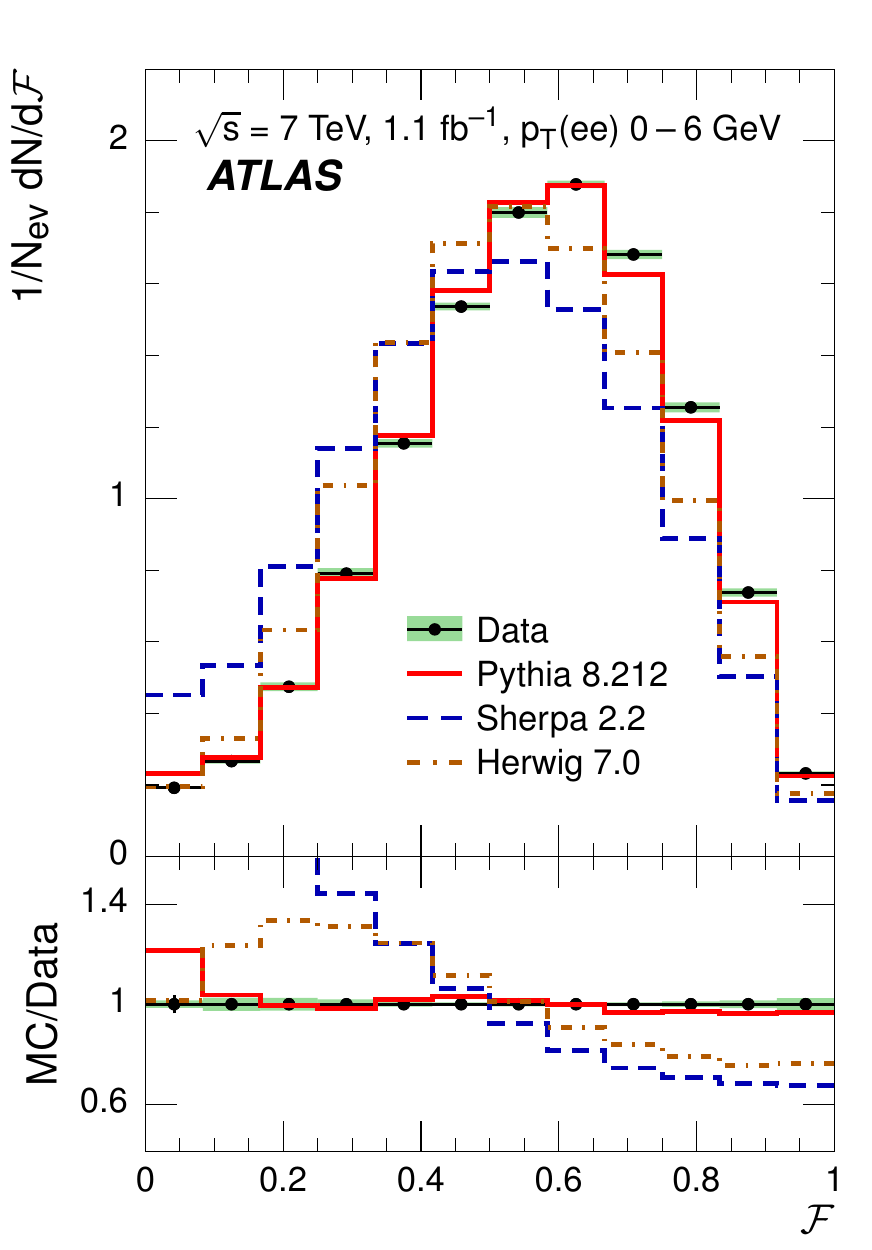}}
(i) \\
\end{minipage}
\hfill
\begin{minipage}[h]{0.24\textwidth} 
\center{\includegraphics[width=1.0\linewidth,height=0.2\textheight]{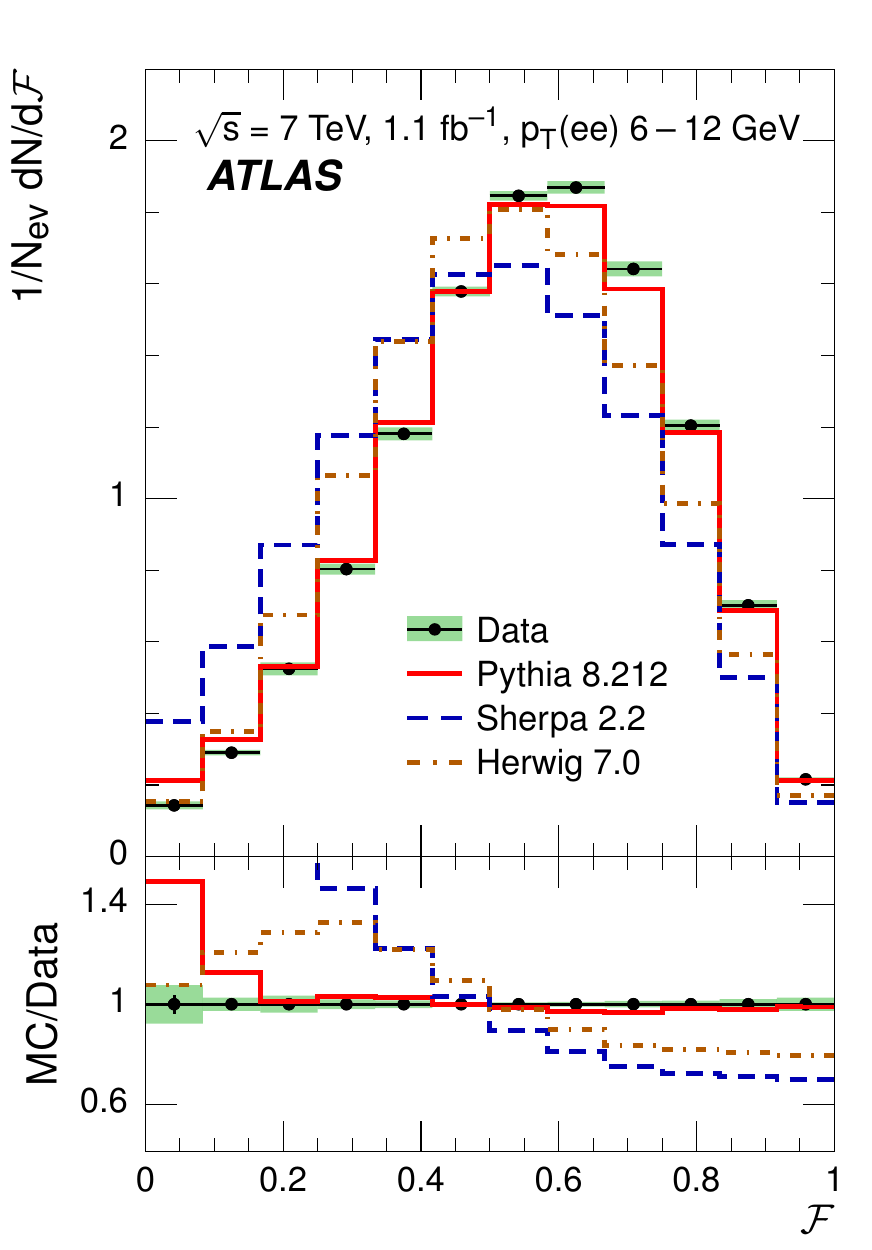}}
(j) \\
\end{minipage}
\hfill
\begin{minipage}[h]{0.24\textwidth}
\center{\includegraphics[width=1\linewidth,height=0.2\textheight]{fig_08b.pdf}} 
(k) \\
\end{minipage}
\hfill
\begin{minipage}[h]{0.24\textwidth} 
\center{\includegraphics[width=1\linewidth,height=0.2\textheight]{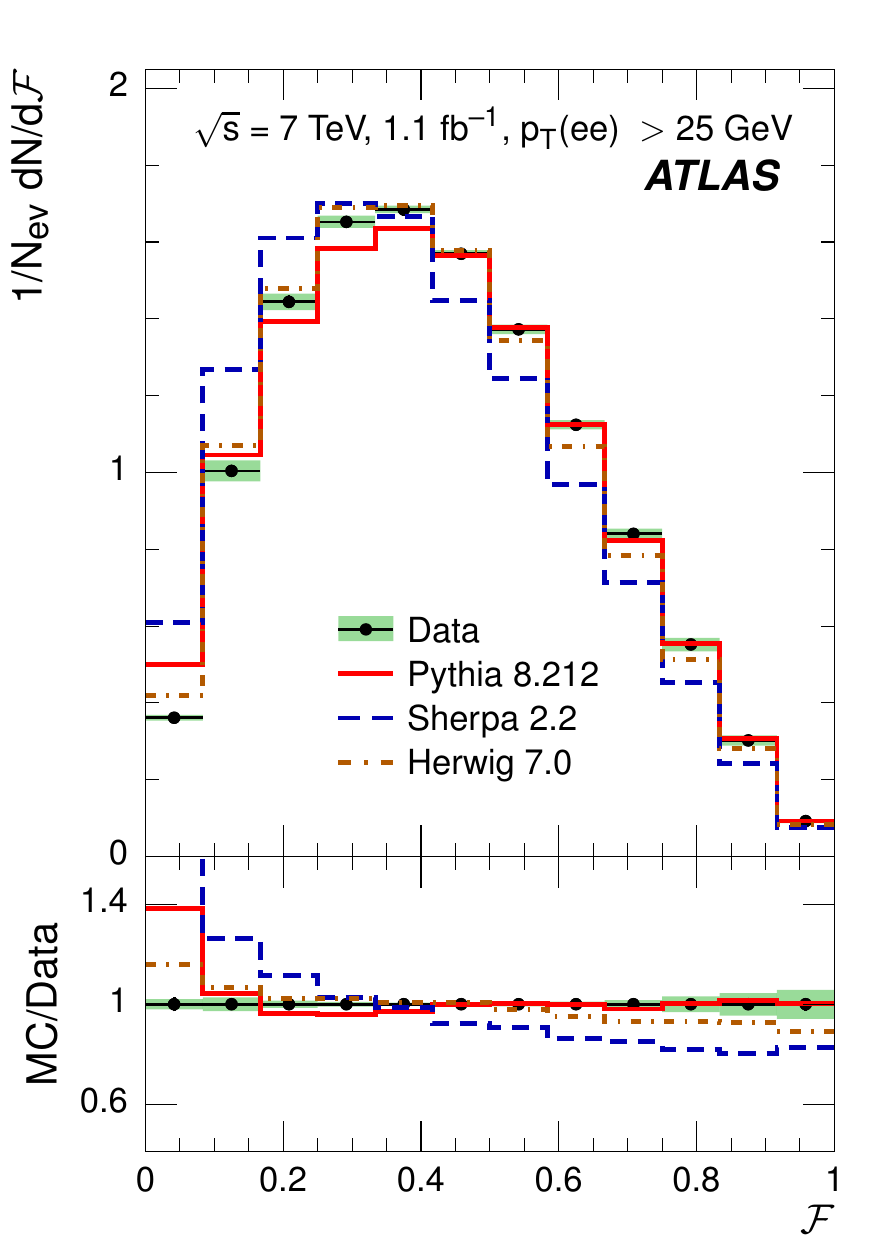}} 
(l) \\
\end{minipage}
\caption{
(a--d) Transverse thrust,
(e--h) spherocity 
and
(i--l)  $F$-parameter 
distributions of charged particles 
for $Z\rightarrow e^+ e^-$ 
for the  $p_{\rm T} (e^+ e^-)$  ranges 
compared to the predictions from the MC generators.
Taken from Ref.\ \cite{Aad:2016ria}.
}
\label{fig_Zll_3}
\end{figure*}
%_______________________________________________________________________

%
The results from the electron and muon channels are in a good agreement.
The Figures show only results for the electron-positron channel.
Figure \ref{fig_Zll_1}  shows the unfolded electron 
channel results for the six observables in the various $p_{\rm T}  (l^+ l^-)$ ranges.
As $p_{\rm T}  (l^+ l^-)$ rises, i.e.\ as recoiling jets emerge, the number of produced charged particles $N_{ch}$ 
increases, as do $\sum p_{\rm T}$ and beam thrust. 
Correspondingly, transverse thrust moves towards higher  values and spherocity 
towards smaller values as a result of the increasing jettiness of the events.
Figures \ref{fig_Zll_2} and  \ref{fig_Zll_3} show the individual event-shape observables for the electron 
channel compared  to predictions obtained with the most recent versions of three different MC generators.
%:
In general, Pythia~8 and Herwig~7 agree better with the data than does Sherpa.
The $p_{\rm T}  (l^+ l^-) < 6$~GeV bin is expected to be characterised by low jet activity from the hard matrix element and
hence should be particularly sensitive to UE characteristics.
In this case, Pythia~8 shows very good agreement with the data in the event-shape observables that are 
not very sensitive to the number of charged particles ($T$, $S$, and $F$-parameter). 
The observables that depend explicitly on the number of charged particles ($N_{ch}$,  $\sum p_{\rm T}$, $B$) are less well
described, with none of the generators succeeding fully. 
In this case, the best agreement is observed for Herwig~7 while Pythia~8 still performs better than Sherpa. 
Low $N_{ch}$ and $\sum p_{\rm T}$ values represent a challenging region for all three generators.
This region might be particularly sensitive to the way beam-remnant interactions are modelled in the MC generators. 
Similar observations can be made for $p_{\rm T}  (l^+ l^-)$ ranges 6--12 and 12--25 GeV. 
At low values of $B$, the observable in which tracks with larger $|\eta|$
values contribute less to the sum of the track transverse momenta, better agreement 
of the generator predictions with the data is observed than at low $\sum p_{\rm T}$.
At $p_{\rm T}  (l^+ l^-) \ge 25$~GeV the event is expected to contain at least one jet of high transverse momentum recoiling
against the Z boson, which is expected to be well described by the hard matrix element. 
In this case, one still observes significant deviations of the MC generators from the measurement, 
where, depending on the observable, either Herwig~7 or Pythia~8 shows in general the best agreement. 
However, all three generators show better agreement with data compared to the $p_{\rm T}  (l^+ l^-) < 6$~GeV range.

%_______________________________________________________________________
%fig 4
\begin{figure*}[t!]
\begin{minipage}[h]{0.24\textwidth}
\center{\includegraphics[width=1.0\linewidth,height=0.2\textheight]{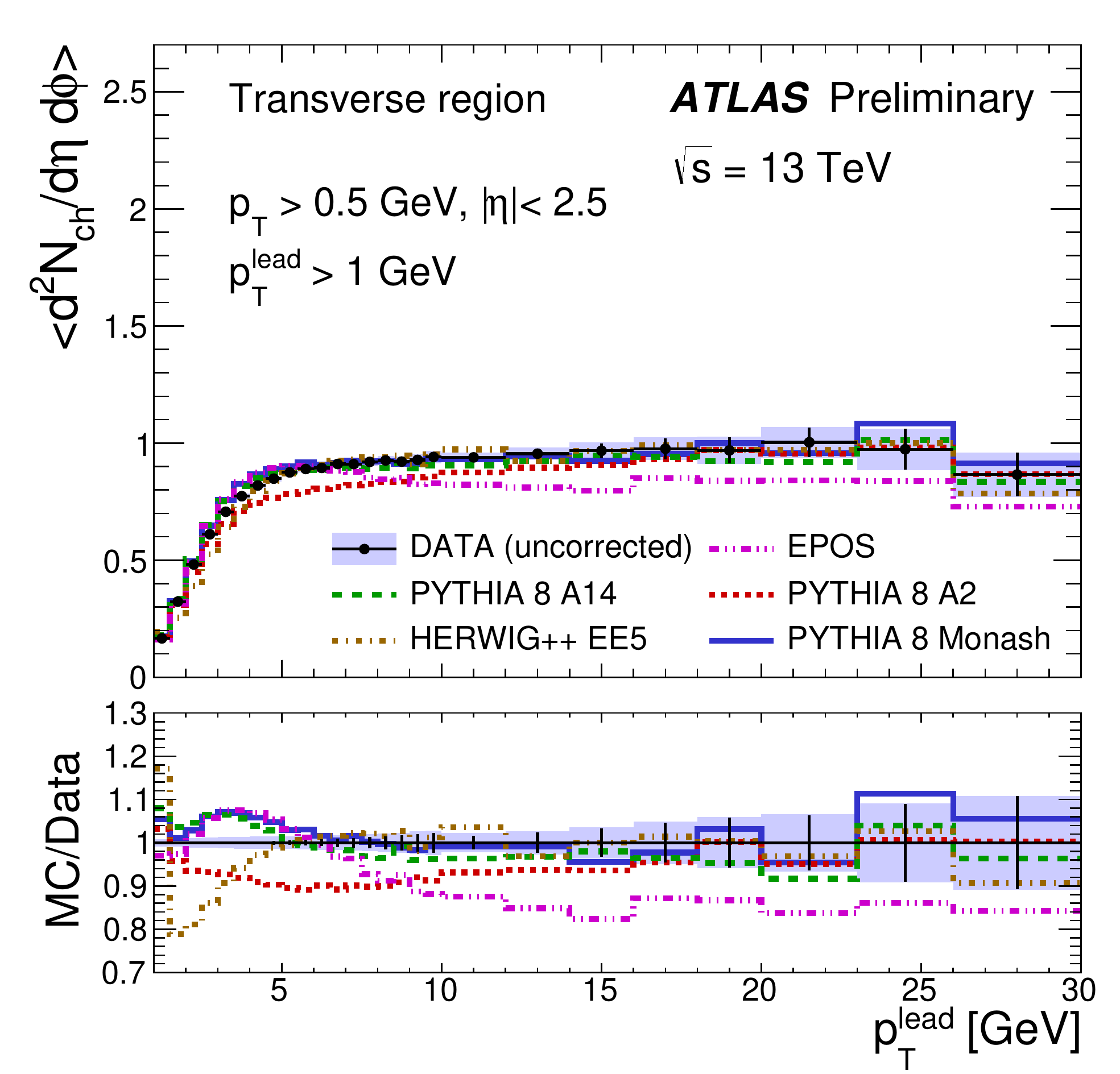}}
(a) \\
\end{minipage}
\hfill
\begin{minipage}[h]{0.24\textwidth} 
\center{\includegraphics[width=1.0\linewidth,height=0.2\textheight]{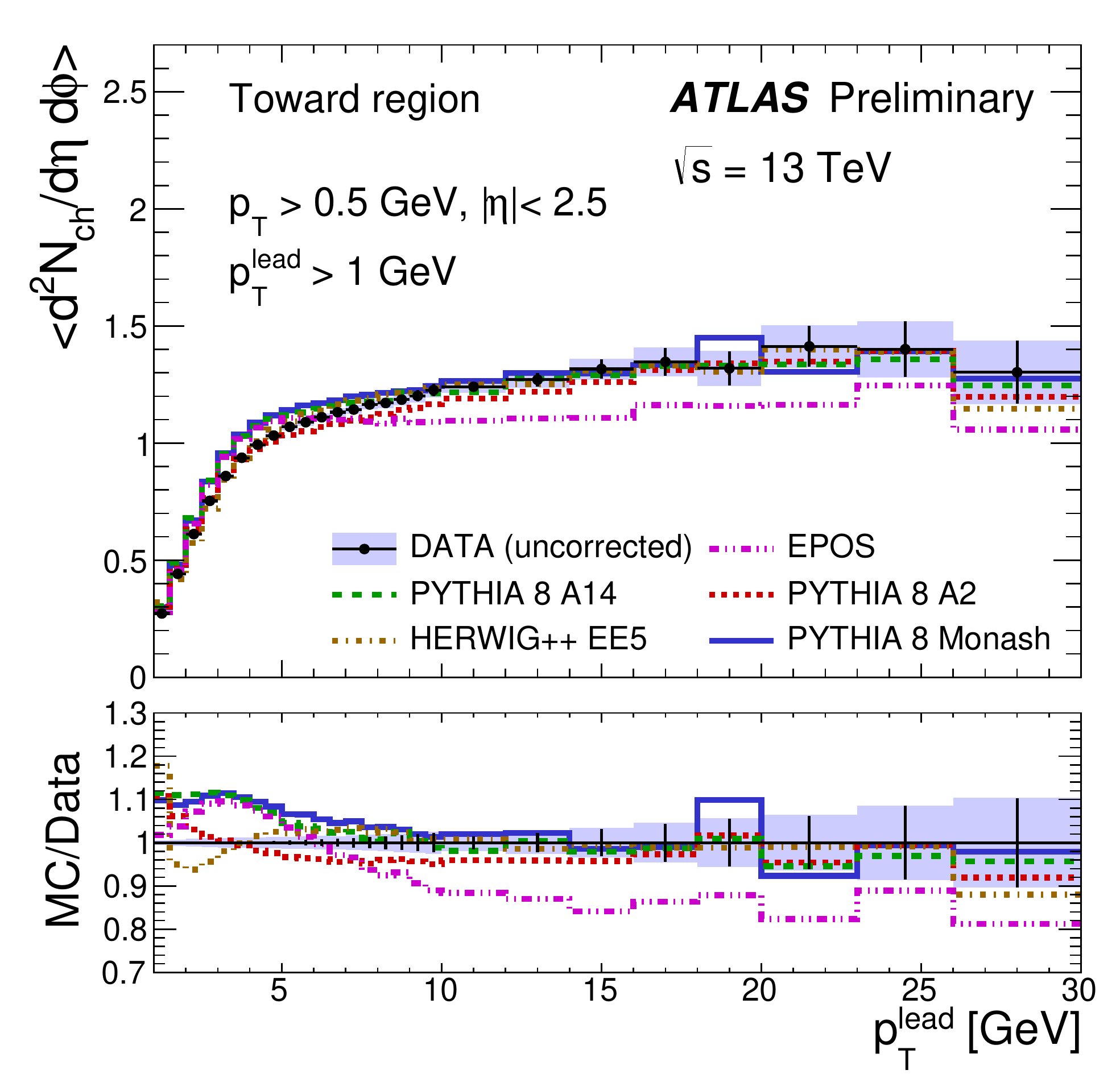}}
(b) \\
\end{minipage}
\hfill
\begin{minipage}[h]{0.24\textwidth}
\center{\includegraphics[width=1\linewidth,height=0.2\textheight]{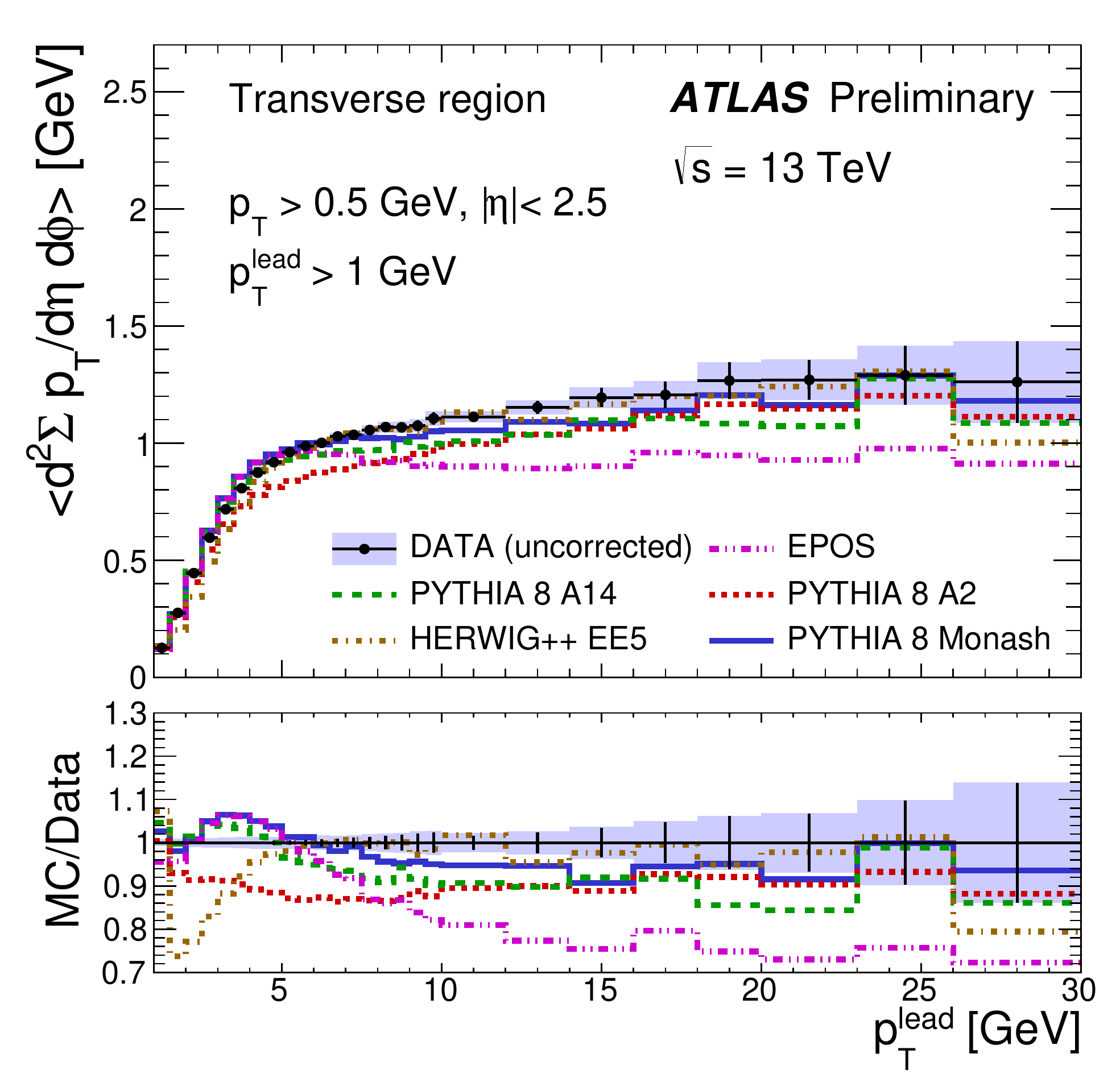}} 
(c) \\
\end{minipage}
\hfill
\begin{minipage}[h]{0.24\textwidth} 
\center{\includegraphics[width=1\linewidth,height=0.2\textheight]{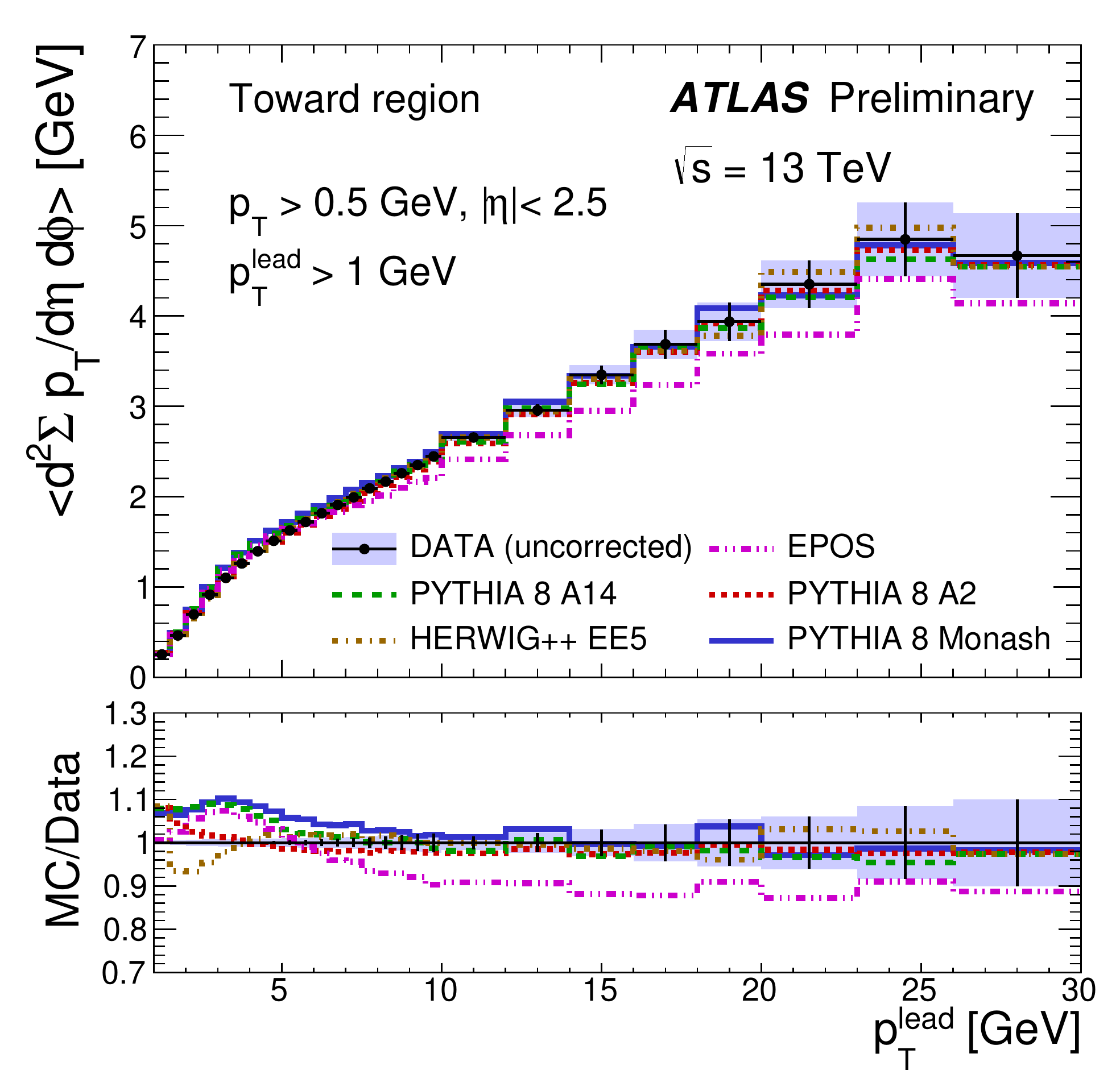}} 
(d) \\
\end{minipage}
\caption{
Comparison of detector-level data and MC predictions for 
 (a, b) 
$\langle d^2 N_{\rm ch}/d\eta d\phi \rangle$ and 
 (c, d)
$\langle d^2 \sum p_{\rm T}/d\eta d\phi \rangle$ 
as a function of 
$p^{\rm lead}_{\rm T}$ in the transverse (a, c) and toward (b, d) regions. 
Taken from Ref.\ \cite{Und_Tr}.
}
\label{fig_udr_1}
\end{figure*}
%_______________________________________________________________________
%

\vspace*{-5mm}
%_______________________________________________________________________
\section{
Leading track underlying event distributions 
}
\label{sec-3}
%_______________________________________________________________________
\vspace*{-2mm}

A detector-level measurement of track distributions sensitive to the properties of the underlying event is 
given in  Ref.\   \cite{Und_Tr}.
It is based on 9 million  events collected using the ATLAS detector  in pp collisions at 
13~TeV, corresponding to an integrated luminosity of  
151~$\mu$b$^{-1}$.
The underlying event (UE) is defined as the activity accompanying any hard scattering in a collision event. 
This includes partons not participating in  the hard-scattering process (beam remnants),
and additional scatters in the same pp collision, termed multiple parton interactions (MPI). 
Initial and final state gluon radiation (ISR, FSR) also contribute to the UE activity. 
The direction of the leading track is used to define regions in the azimuthal plane that have different
sensitivity to the UE.
The direction of the leading track is used to define regions in the azimuthal plane that have different
sensitivity to the UE, a concept first used in \cite{CDF}. 
The azimuthal angular difference between tracks and the leading track 
$|\Delta\phi|= |\phi- \phi^{lead}_{track}|$, is used to define the following three azimuthal UE regions: 
1) $|\Delta\phi| < 60^\circ$, the toward region,
2) $60^\circ <|\Delta\phi| < 120^\circ$, the transverse region, 
and
3) $|\Delta\phi| > 120^\circ$, the away region.
The transverse region is sensitive to the underlying event, since it is by construction perpendicular
to the direction of the hard scatter and hence it is expected to have a lower level of activity from the
hard scattering process compared to the away region. 
The observables measured in this analysis are 
%1) 
number of tracks per unit of $\eta$--$\phi$, $\langle d^2 N_{\rm ch}/d\eta d\phi \rangle$; 
and 
%2) 
scalar sum of track $p_{\rm T}$ per unit of $\eta$--$\phi$, $\langle d^2 \sum p_{\rm T}/d\eta d\phi \rangle$.
The average density of particles, $\langle d^2 N_{\rm ch}/d\eta d\phi \rangle$, 
and average transverse momentum sum density, $\langle d^2 \sum p_{\rm T}/d\eta d\phi \rangle$,
are constructed by dividing the mean values by the corresponding area. 
The leading track is included in the toward region distributions, unless otherwise stated.
The results are presented at detector-level, without any corrections (except in MC, the width of the
vertex distribution along the $Z$ axis is reweighted to match the data). 
The uncertainty resulting from tracking inefficiency is estimated by randomly dropping tracks according 
to the tracking efficiency uncertainty determined as a function of track $p_{\rm T}$ and $\eta$. 
In Fig.\ \ref{fig_udr_1} the average densities of track multiplicity and scalar $\sum p_{\rm T}$ are shown. 
In the transverse region, both show a gradual increase, rising to an approximately constant “plateau” 
for $p^{\rm lead}_{\rm T} > 6$ GeV. 
For higher values of $p^{\rm lead}_{\rm T}$, the toward and away regions include contributions from jet-like activity, 
yielding gradually rising 
$\langle \sum p_{\rm T} \rangle$ density. 
Among the MC models, Pythia~8  A14, Monash \cite{UE_shape_2}  and Herwig++  \cite{UE_shape_1}
predictions are closest to data in the plateau  part of the transverse region, but none of the models describe the initial rise well. 
The EPOS \cite{epos}  generator predicts significantly less activity at higher $p^{\rm lead}_{\rm T}$, 
indicating the absence of semi-hard MB events. 

\vspace*{-5mm}
%_______________________________________________________________________
\section{Summary}
%_______________________________________________________________________
\vspace*{-2mm}

 A correct modelling of the underlying event in  pp-collisions is important for the proper 
 simulation of kinematic distributions of  high-energy collisions.
Event-shape observables sensitive to the underlying event were measured in 1.1~fb$^{-1}$ integrated luminosity of pp-collisions collected with the ATLAS detector at the LHC at a centre-of-mass energy of 7~TeV.
 The distributions of several topological event-shape variables based on charged particles are measured
  for events containing an oppositely charged electron or muon pair with an invariant mass close to the Z-boson mass.
 The collaboration has also performed a first study of the number and  transverse momentum
 sum of charged particles as a function of  transverse momentum and azimuthal angle in a special 
 data set taken with low beam currents with integrated luminosity of  
 151~$\mu$b$^{-1}$ at a  center-of-mass energy of 13~TeV. 
 The results are compared to predictions of several MC generators. 
These are required  for tuning of the soft  part of Monte Carlo simulation.

\vspace*{-5mm}
%_______________________________________________________________________
{%\footnotesize

}
%_______________________________________________________________________
 
%_______________________________________________________________________
\end{document}